\documentclass[proof]{WileyASNA-v1}

\articletype{Article Type}%
\usepackage{apacite}
\revised{27 Sep 2023}
\accepted{29 Jan 2024}

\usepackage{graphicx}   
\usepackage{amsmath}    
\usepackage{amssymb}    
\usepackage{longtable}
\usepackage{textgreek}
\usepackage{multicol}
\usepackage{lscape}
\usepackage[T1]{fontenc}
\usepackage{ae,aecompl}
\usepackage[LGR,T1]{fontenc} 
\usepackage{longtable}

\DeclareSymbolFont{greekletters}{LGR}{\familydefault}{m}{n}
\DeclareMathAccent{\dasia}{\mathord}{greekletters}{60}
\DeclareMathAccent{\psili}{\mathord}{greekletters}{62}

\begin{document}

\sloppy

\title{The Mira discovery problem -- Observations by David Fabricius in 1596 and 1609 (and by others before?):
Positional accuracy, brightness, color index, and period}

\author[1]{R. Neuh\"auser (*)}
\author[2]{D.L. Neuh\"auser (*)}
\author[1]{M. Mugrauer}
\author[1]{D. Luge}
\author[3]{J. Chapman}

\authormark{R. Neuh\"auser \& D.L. Neuh\"auser \textsc{et al.} Observations of Mira by Fabricius}

\address[1]{\orgdiv{Astrophysikalisches Institut}, \orgname{Universit\"at Jena}, \orgaddress{\state{Schillerg\"a\ss chen 2-3, 07745 Jena}, \country{Germany}}}
\address[2]{Independent scholar, Merano, Italy}
\address[3]{\orgdiv{Institute of East Asian Studies}, \orgaddress{UC Berkeley, Berkeley CA, 94720}, \country{USA}}

\corres{*Corresponding author name: Ralph Neuh\"auser. \email{ralph.neuhaeuser@uni-jena.de}}
\corres{*These authors contributed equally. Corresponding author: Ralph Neuh\"auser. \email{ralph.neuhaeuser@uni-jena.de}}

\abstract{The pulsating variable star Mira (${\it o}$ Ceti) was observed by David Fabricius (Frisia) in 1596 and 1609.
We review suggested previous detections (e.g. China, Hipparchos). 
We analyze all Mira records from Fabricius in their historical context. 
Fabricius measured the separation of Mira to other stars to $\pm$1.6-1.7$^{\prime}$.
From his texts, we derive a brightness (slightly brighter than Hamal) of $\sim$1.9$\pm$0.1 mag 
and a color index B-V$\simeq$1.3-1.4 mag 
(`like Mars') for 1596 Aug 3 {(jul.)}. Mira started to fainten 19 days later and was observed until mid/late Oct.
We show why such a red star cannot be followed by the naked eye
until $\sim$6 mag: For Mira's color at disappearance and altitude from Frisia,
the limit is reduced by $\sim$1.0 mag.
Since Fabricius connected the Mira brightening with the close-by prograde Jupiter, he re-detected it only 12 years later, 
probably shortly before a relatively bright maximum -- discoveries are strongly affected by biases.
A Mira period of 330.2 days is consistent with both the oldest data (from Fabricius 1596 to Hevelius 1660)
and the most current data (VSX 2004-2023), so that we see no evidence for secular period or phase shifts.
(We also present Fabricius' observations of P Cygni in 1602.)}


\maketitle


\section{Introduction}

This study is part of a larger project to compile pre-telescopic (i.e. naked eye) records on star colors
in order (a) to derive the brightness limit for naked-eye color detection of stars in an empirical way
and (b) to consider possible secular color changes over centuries and millennia due to stellar evolution.
Previously, we published the rapid color evolution of Betelgeuse ($\alpha$ Ori) from B$-$V$\simeq$1.0 mag
some two millennia ago to its current deep red color (B$-$V=$1.78 \pm 0.05$ mag) -- based on
several independent historical records from the Mediterranean (Hyginus, Ptolemy, etc.) and China (Sima Qian)
as well as theoretical evolutionary model tracks (MESA), see R. Neuh\"auser et al. (2022).
In a forthcoming publication, we plan to present and discuss all other pre-telescopic color records found
(R. Neuh\"auser \& D.L. Neuh\"auser et al., in prep.).

Here, we concentrate on Mira\footnote{Named Mira from the Latin `mira' for `wonderful' by J. Hevelius in his
work on this star entitled `Historiola Mirae Stellae' (Hevelius 1662); this naming may have been motivated by the fact
that, when Fabricius wrote to Kepler on his observations of this star in 1609, he called the story twice
`res mira', i.e. `wonderful thing' (Bunte 1888, p. 5).} (omicron Ceti or {\it o} Ceti\footnote{Bayer (1603)}) and its discovery. 
Given the evolutionary stage of the pulsating giant stars, Mira (see Whitelock 1999 for a review)
is another example, where an astrophysical process can possibly be studied on the historical 
time scale ({\it Terra-Astronomy} or Applied Historical Astronomy, 
see R. Neuh\"auser, D.L. Neuh\"auser, Posch 2020), 
here stellar evolution (e.g., secular period or phase shifts in Mira).

In this paper, we present (for the first time in astronomical literature) all known historical contemporaneous 
records on the discovery of Mira 
by David Fabricius in 1596 and his re-detection in 1609 -- both in the original version (partly Latin, partly old German) 
and our English translations. The texts are from Fabricius himself and partly citations
from Brahe and Kepler of letters from Fabricius. (Two of these texts are also found in Hatch (2011) in his essay
on `Discovering Mira Ceti', but his work is mainly on the extensive observations by Hevelius and Boulliau much later.)

David Fabricius was born on 1564 Mar 9 in Esens (present-day northern Germany) to Talke and Jan Jansen,
was immatriculated as `David Faber Esenesis' to study theology 1583-1584 at U Helmstedt, Germany,
then worked as protestant pastor first in Resterhafe and since 1604 in Osteel, both East Frisia in northern Germany,
where he was murdered on 1617 May 7 (see Folkerts 1997).
The name `Fabricius' is the Latin form of faber or goldsmith, the profession of his father.
David Fabricius and his oldest son, Johann (born 1587), are known also for having observed the Sun already
in Feb 1611 with a telescope, whereby they detected spots and published a first paper 
about those spots and the solar rotation period (see e.g. R. Neuh\"auser \& D.L. Neuh\"auser 2016).
David Fabricius also observed the supernova of 1604 with important data
for the light curve (Baade 1943) and the most precise position, $\pm 1^{\prime}$ (Schoenfeld 1865, Schlier 1934, Baade 1943),
so that Green (2005) wrote that naming ``{\it V843 Oph} [=SN 1604] {\it as `Kepler's is rather misinformed}''.
Fabricius has built by himself astronomical instruments including a quadrant, a sextant,
and a visier instrument. 

We analyse here mainly the observations of David Fabricius on Mira in 1596
regarding position, brightness (maximum 2nd mag), and color.
The 1596 maximum might well be one of the brightest maxima of Mira in more than four centuries.
We derive the B$-$V color index for the 1596 maximum 
from both the historical transmission and
in comparison
to modern light curves of Mira in B and V. 
His observations in 1609 suffered from the upcoming conjunction with the Sun,
and he did not provide brightness or color information.

We focus on the Mira discovery problem: 
Why was it not detected more often in pre-telescopic time, since it can reach ca. 2nd mag at maximum?
Mira is a pulsating variable star, whose brightness (ca. 2nd-10th mag) and color change
significantly with a period of about 331 days.
While David Fabricius reported Mira to be slightly brighter than $\alpha$ Ari (ca. 2 mag) 
for the discovery on 1596 Aug 3 (Julian calendar, henceforth {\it jul.}, corresponding to Aug 13 on
the Gregorian calendar, {\it greg.}), he followed it only
until October 1596 -- according to modern light curves, this would not be down to ca. 6th mag,
the conventional naked-eye limit.
We discuss the reason for why Fabricius lost Mira when still being significantly brighter than 6th mag --
partly in comparison to modern naked-eye and CCD observations as well as data from Hevelius and Argelander.

It is questionable whether Mira was detected earlier -- as part of a constellation
(e.g. Manitius 1894 for Hipparchos) and/or as guest star in China (AD 1070) and/or Korea (AD 1592). 
Since Mira reaches 2nd to 4th mag in the maxima, Clark \& Stephenson (1977, pp. 50-51)
concluded from the presumably rare (or missing) detections of Mira that serendipitously discovered 
new stars (supernovae, novae, etc.) are typically brighter than 3rd mag -- and, given typical
supernova light curves, that supernovae are then visible for at least 40 days to reach 5.5 mag.
We reflect anew on the question of serendipitous discovery of variable stars and on why
Fabricius needed 12 years to detect Mira again.

In this paper, we first discuss the original sources for the possible observations of Mira before 
Fabricius by Hipparchos as well as in China and Korea (Sect. 2).
Then, we present the original texts from Fabricius, Brahe, and Kepler for 1596 and 1609
with our English translations and analyse them in the historical context (Sect. 3).
We discuss the positional accuracy of Fabricius' data as well as brightness and color index of Mira 
during the bright maximum of 1596 in Sect. 4 and derive astrophysical conclusions
(also regarding possible period and/or phase shifts). 
We finish with a summary in Sect. 5.

\section{Possible observations of Mira before Fabricius (1596)}

We know of a few suggestions regarding observations of Mira before Fabricius in 1596,
which we discuss here.

{\bf (i) Was Mira a constituent of Hipparchos' constellation Cetus?} \\
M\"uller \& Hartwig (1920, p. 449) noticed a possible early record on Mira by Hipparchos
(flourished BC 147-127, probably on Rhodos, Greece):
Based on Manitius' edition and translation from the original Greek to German (Manitius 1894)
of Hipparchos' commentary on Aratos and Eudoxos,
{\it o} Ceti may have been part of the constellation Cetus --
here the relevant lines (round brackets from Manitius, square brackets from us),
our literal translation from German:
\begin{quotation}
\noindent ... Cetus ... it does not rise together with Aries, as they claim, but with the Pisces,
nor does it set until the backfin [lophias] ({\it o}) of Cetus [Manitius, pp. 182 and 183, line 3] \\
... of Cetus on its backfin [lophias] ({\it o}) [pp. 264 and 265, line 17]. \\
... the knot ($\alpha$) in the band of Pisces, which stands in the area of the head
of Cetus on its backfin [lophias], stands on the meridian of Aries 2.25$^{\circ}$ [pp. 120 and 121, line 20], \\
The remaining body of Cetus does not set fully with the rise of Virgo,
but only up to its backfin [gr.: lophias], as Aratos writes [pp. 160 and 161, line 11], \\
\end{quotation}
Manitius identifies the star on the backfin (lophia) of Cetus twice explicitly with {\it o} Ceti (Mira).
Manitius did not give this identification ({\it o}) for two other lines (Manitius 1894, pp. 120-121 and 160-161), 
where Hipparchos also had {\it lophia} in respect to Cetus.
The relevant Greek term {\it lophia} mainly means `mane' (e.g. of a horse).

Ptolemy did not use {\it lophia} at all in the Almagest for any stars in Cetus.
However, a star `about on the mane' close to the head of Cetus is mentioned,
i.e. Cetus 7 = $\xi^{1}$ Ceti (Toomer 1984);
Ptolemy gave here a different Greek word, namely {\it chait\=e}.
The star $\xi^{1}$ Ceti is a certain identification (see Toomer 1984, pp. 381-382).
In addition, there is a star located `about on the hair', 
credibly identified as Cetus 6 = $\xi^{2}$ Ceti by Toomer (1984);
Ptolemy used here {\it thrix}.
According to Toomer (1984), it is clearly distinct from the `mane' of the star Cetus 7.

Manitius (1912) gave a translation of the Greek Almagest to German
with the same wording (German: `M\"ahne' (mane) for Cetus 7 and `(Stirn-)Haar' (front-hair) for Cetus 6.

It seems that Manitius thought that the different word {\it lophia} used by Hipparchos would
then pertain to another (different) star.
The Greek term {\it lophia} (`mane') is related to {\it lophos}, which means among others `neck'.
So, he then picked {\it o} Ceti, which he noticed in the catalog by Bayer (1603),
located there on the neck of Cetus (Bayer: `next to the curvature or hump') --
Manitius mentioned Bayer's work explicitly in his introduction.
That Manitius translated and interpreted it as `R\"uckenflosse' (`backfin')
may be influenced by recent translations of {\it lophia} (see Liddell \& Scott 1843) and/or depictions of Cetus.
A star on the backfin of Cetus is otherwise not known, e.g. in the Almagest.

In the work by Hipparchos, the terms {\it chait\=e} and {\it thrix} do not appear (Manitius 1894),
but (only) {\it lophia} (`mane'), so that we conclude that 
Hipparchos' `lophia' star is $\xi^{1}$ Ceti (or maybe $\xi^{2}$ Ceti).

The rising and setting constrains given by Hipparchos -- as cited above --
for the star on the mane (lophia), identified by Manitius as Mira, holds for $\xi^{1}$ Ceti.
Since Hipparchos is interested in improving rising and setting data of stars and the
constellation patterns by Aratos and Eudoxos, it would not be convincing to include
a star that is not visible for more than half of the time.

N.B.: In Ptolemy's Almagest, Mira is apparently not included (e.g. Manitius 1912, Toomer 1984).
The first (and only) pre-telescopic star catalog {\it Uranometria} listing Mira explicitly is the one by Bayer (1603, 
Johann Bayer was born 1572 in Rain, studied philosophy and law at U Ingolstadt,
worked as lawyer for the city of Augsburg, and died in 1625, all in present-day Germany).
We consulted the Uranometria manuscript ETH Z\"urich Rar 8931 q, 
where Bayer provided drawings of the constellations with stars plus text: \\
`24 -- {\it o} [Ceti] -- iuxta curvaturam seu gibbum -- quartae', i.e. \\
`24 -- {\it o} [Ceti] -- next to the curvature or hump -- 4th [magnitude]'. \\
For Bayer, it is the 24th star in Cetus, while Ptolemy had 22 in total, i.e. one of the added stars. 
In the drawing, {\it o} Ceti is placed correctly in the lower neck.\footnote{There
are some peculiarities in Bayer's catalog:
star no. 5 in the Almagest has no. 25 in Bayer, and the Almagest stars Cetus 17, 18, and 19 
are 3 of the 4 stars summarized by Bayer under his entry no. 20.
See Ridpath (1988) for more details.}
The star in the mane is $\xi^{1}$ Ceti (like in the Almagest). 
Hevelius, in his star catalog, described Mira to be located at the neck (lat. collum) of Cetus.
Bayer's source for Mira is not clear, e.g., whether he observed it himself and/or 
whether he knew of Fabricius' observations; the 4th mag given by Bayer is not found in
the text we have from Fabricius (Sect. 3).

\smallskip

Then, M\"uller \& Hartwig (1920, p. 449) have an additional argument:
They interpret the short text from Pliny (AD 23/24 to 79, Roman lawyer and author)
on the `new star' of Hipparchos\footnote{Pliny (Nat. Hist. II, 24):  
`Hipparchos ... detected a new star [lat. stella nova] that came into existence during his lifetime; 
the movement of this star in its line of radiance 
led him to wonder whether this was a frequent occurrence, whether the stars that
we think to be fixed are also in motion; and consequently he did a bold thing ... he
dared to schedule the stars for posterity, and tick off the heavenly bodies by name
in a list, devising machinery [armillary sphere?] by means of which to indicate their several positions
and magnitudes, in order that from that time onward it might be possible easily to
discern not only whether stars perish and are born, but whether some are in transit
and in motion, and also whether they increase and decrease in magnitude'
(cited after Rackham et al. 1938). \\
If the object moved relative to the stars (Pliny: `movement' and `motion'),
it might have been a comet; during the observing period of Hipparchos 
as mentioned in the Almagest (BC 147-127), there were at least two bright comets,
namely in BC 135 
and BC 147 (Kronk 1999, Pankenier 2013);
it is credible that Hipparchos may have tried to
record the path of a comet in comparison to certain stars, so that he measured their positions and
brightness relative to others in their surrounding. \\
Future studies (`from that time onward') could then address three problems:
(i) `whether stars perish and are born', which may indicate that new or disappearing stars 
(in today's sense, e.g., novae or supernovae) were already observed,
(ii) to study the path of comets (`whether some are in transit and in motion'), 
including their possible periodicity, which was considered already in antiquity, 
(iii) brightness variability of stars (`increase and decrease in magnitude'),
which may imply that such phenonema like in Algol or Mira were already observed.
Thereby, Hipparchos could have been motivated to compile a star catalog.}
such that Hipparchos would have discovered {\it two} new stars, one of them {\it o} Ceti following Manitius (1894), 
which is not justified as shown above;
for the presumably second new star, M\"uller \& Hartwig (1920, p. 449) cite Manitius (1894, p. 203) 
for a star in Cetus with `der südlich davon [$\tau$ Ceti] stehende unbekannte helle', i.e.
`the unknown bright one standing to the south of it [of $\tau$ Ceti]';
however, Manitius did not write `unbekannte' (English: unknown), but `unbenannte' (English: unnamed, nameless),
the correct translation of the Greek `akatonómastos' in Hipparchos' work,
twice found in Manitius (1894, pp. 203 and 247).
(Hipparchos wanted to `tick off the heavenly bodies by name in a list', see footnote 4.)
There is no evidence for a second new star.

\bigskip

{\bf (ii) Guest star AD 1592-1594 observed in Korea:} \\
Brosche (1966) suggested a guest star reported from Korea for AD 1592 to 1594 in Tiancang as possible early
detection of Mira. Clark \& Stephenson (1977) followed suit.
One of the reports from Korea (Yijo sillok Sonjo ch. 31-47), as translated by Xu et al. (2000), is 
as follows:
\begin{quotation}
\noindent the guest star was at Tiancang, within the third star from the east and about three cun away.
Its color and form were less (conspicuous) than the stars of Tiancang.
\end{quotation}
The text in Chinese characters is found in Xu et al. (2000, p. 337).
There are several such reports in that chronicle from 1592 Nov 23 to 1594 Feb 23 ({\it greg.}),
discussed also in Stephenson \& Yau (1987) and Huang (1988).

A separation of `about 3 cun' is less than half a degree 
(1 cun = 0.1 chi = $0.15 \pm 0.24^{\circ}$ according to Kiang 1972) 
off the 3rd star of Tiancang 
from the east, which is $\zeta$ Ceti according to, e.g., Sun \& Kistemaker (1997), see also our Fig. 1. 

Even though the observations 1592-1594 may be roughly in phase with the variability of Mira
(when extrapolating back from Fabricius 1596), Hoffleit (1997) 
argued that the long visibility duration of roughly 7 months 
(maybe even longer, if uninterrupted) 
is too long given the typical Mira light curve. Furthermore,
this guest star cannot be Mira, 
because the given position close to $\zeta$ Ceti is strongly inconsistent with Mira, and
there is no other visible star within `3 cun' ($\le$0.5$^{\circ}$) of Mira (e.g., Stephenson \& Yau 1987).

It may be surprising that the observations of the above guest star since 1592 Nov 23
started in the same lunar month 20 years after the AD 1572 
supernova record and lasted also 15 months like the SN 1572 observation in China. However, since the details of the records
are different (e.g. location, color, etc.), the 1592 records are not mis-dated copies of the 1572 event.

\begin{figure*}
\centering
\includegraphics[angle=0,width=7cm]{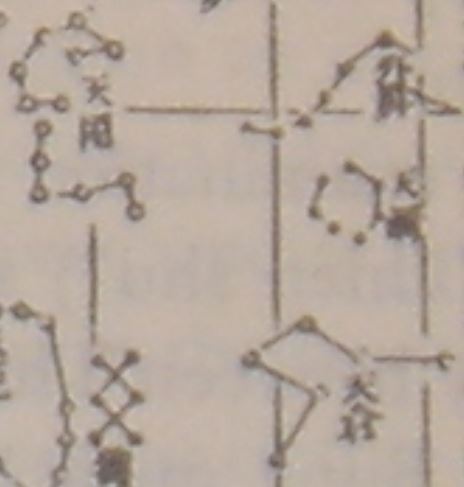}
\hspace{.3cm}
\includegraphics[angle=0,width=8.5cm]{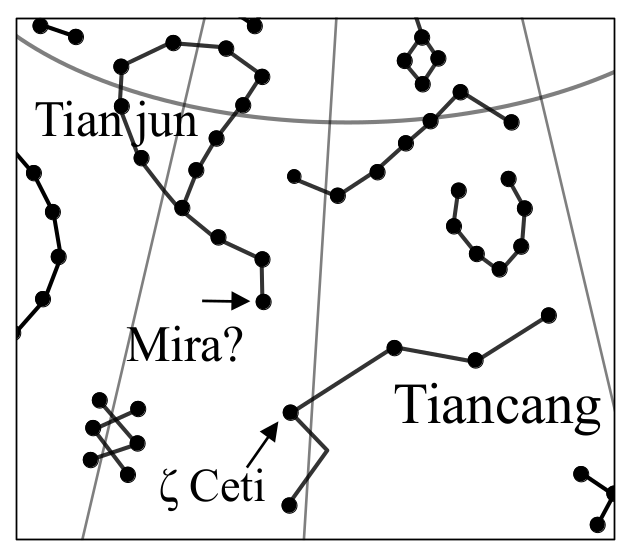}

\vspace{.3cm}

\mbox{\includegraphics[angle=0,width=8cm]{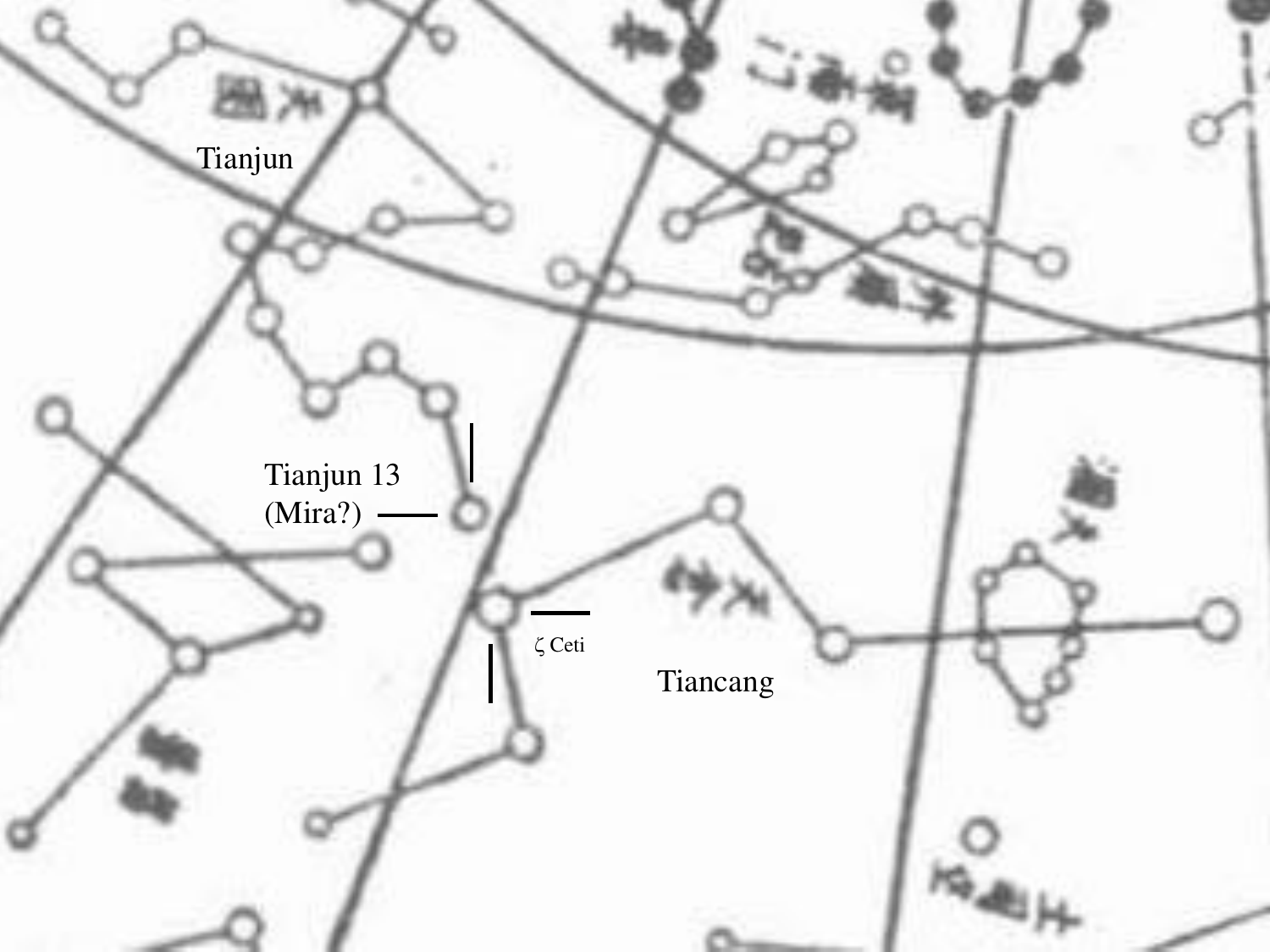}}
\hspace{.3cm}
\includegraphics[angle=0,width=8cm]{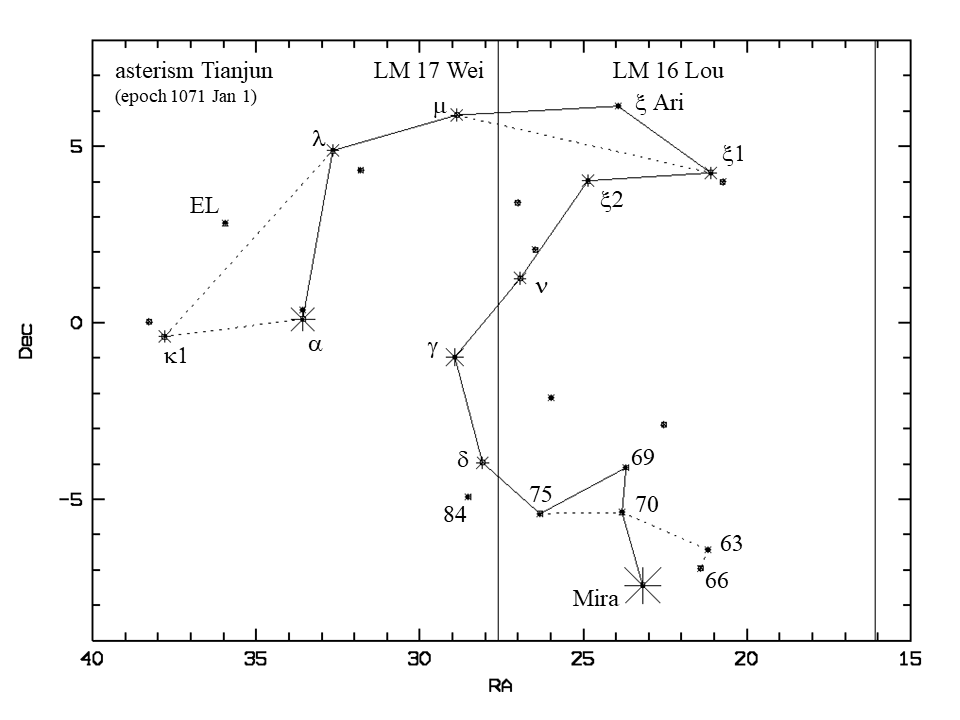}
\caption{{\bf Is Mira a constituent star of Tianjun as mapped on ancient China star charts?} \\
{\bf Upper left:} A small part of the `Xin Yi Xiang Fa Yao' star map from
Su Song (AD 1020-1101) roughly in equidistant, quasi-orthomorphic cylindrical (Mercator) projection 
(public domain, also shown in Needham \& Wang 1959, p. 277), but somewhat abstracted (rectangular), 
with the asterism Tianjun in the upper left (roughly NE), where Mira might be the last star in the south,
and the asterism Tiancang in the lower right with $\zeta$ Ceti as `3rd star from the East' (Sect. 2\,ii).
The lines indicate the celestial equator and borders of lunar mansions. \\
{\bf Upper right:} Small part of a caisson ceiling piece of ca. AD 1453, Longfu Temple, Beijing;
redrawn by us based on figure 2 in Morgan (2018, see \url{https://shs.hal.science/halshs-01714768}), 
where it was taken from Chen Meidong (1996, plate 7-2); we indicate Mira and $\zeta$ Ceti;
a photograph and a drawing of this map are also shown in Stephenson (1994);
the straight lines are again lunar mansion borders, the curved one the equator. \\
{\bf Lower left:}
Small part of the Ming paper planisphere from after AD 1572 (SN shown) and before AD 1644 (end of Ming),
public domain, also shown in Stephenson (1994); we indicate Mira and $\zeta$ Ceti. 
Tianjun is depicted in the upper left, where Mira might be the last star in the south,
and Tiancang in the lower right with $\zeta$ Ceti as `3rd star from the East'.
The lines indicate the celestial equator, the ecliptic, and borders of lunar mansions. \\
{\bf Lower Right:} We show the modern reconstructions of Tianjun from Sun \& Kistemaker (1997, full lines)
and from Yi Shitong (1984, variations as dotted lines ending with 66 Ceti).
Equatorial coordinates right ascension and declination for epoch 1071 Jan 1; 
symbol sizes for stars relate to their brightness (smallest for V=5-6 mag, Mira for V=2 mag); 
all stars are in Cetus (except $\xi$ Ari); all stars near the skeleton lines brighter than V=6 mag are shown; 
the full vertical lines indicate the western borders of the lunar mansions Lou and Wei at $\beta$ Ari and 35 Ari.
}
\end{figure*}

\bigskip

{\bf (iii) Guest star in AD 1070 observed in China:} \\
Clark \& Stephenson (1977) mentioned a guest star reported for
AD 1070 Dec 25 in China as an additional possibility for Mira, also Hoffleit (1997):
\begin{quotation}
\noindent Emperor Shenzong, 3rd year of the Xining reign period, 11th month, day dingwei [44].
A guest star emerged at Tianqun [Tianjun]
\end{quotation}
from Song shi `Tianwen zhi' ch. 56 as translated by Xu et al. (2000).
The text in Chinese characters is found in Xu et al. (2000, p. 334).

A phenomenon called `guest star', but given without any explicit duration (maybe just for or within
one night) could be any kind of phenomenon including a comet, meteor, bolide, or fast nova --
the latter was suggested by Hsi (1957). 
The constellation is transcribed as {\it Tianqun} in Xu et al. (2000) and in table 3.1 by Sun \& Kistemaker (1997, p. 51),
and as {\it Tianjun} in Sun \& Kistemaker (1997) on their chart and in their index,
which are both given in various dictionaries;
Stephenson \& Green (2009) have {\it Tianchun}, which seems to be an otherwise unused combination of the first term 
from the current Pinyin system (tian-jun/-cun) and the 2nd term from the old Wade-Giles transliteration (t'ien-ch'un).

Mira indeed lies in the area of this asterism (Fig. 1), so that it could be a possibility.

Stephenson \& Green (2009) mention {\it Xu Zizhi tongjian changbian} 
(`Extended Continuation to the Comprehensive Mirror that Aids in Government', 
compiled by Li Tao in AD 1183, chap. 217)
as a second source for AD 1070/71, but quote only the phrase `the guest star appeared at Lou'
to be given for the same `dingwei' day (44) in month 11 of the same year, AD 1070 Dec 25. 

We have consulted this source and found the following text:
\begin{quotation}
\noindent Dingwei, kexing chu Lou. Xin, jiu ji xin shilu chao shu ci. 
Tian wen zhi yun: Kexing chu Tianjun xing zhong, zhu cang ku you huo zai.
\end{quotation}
The text in Chinese characters is found in the Appendix.
NB: True (or veritable) records chronicle the major affairs occurring during the reign of a single emperor. 
The closest analogue to this quotation in an extant {\it Treatise on Celestial Patterns} from the standard histories 
occurs in Song shi (History of the Song Dynasty) 56.1230 (also discussed here), but it must be noted that the Song shi 
was compiled by Toghto (Tuotuo, 1313–1355) from 1343–1345, long after the completion of the Xu Zizhi tongjian changbian 
in 1183. The record in the Song shi gives the same date, but does not mention the danger of fires in silos and granaries.

Our literal translation is as follows (our additions in brackets):
\begin{quotation}
\noindent [Day] dingwei (44), [a] guest star emerged [in/at] Lou.  
New [and] old annals followed true records [by] copying [and] recording this.  
`Tianwen zhi' (i.e. `Treatise on Celestial Patterns') says: 
[When a] guest star emerges in the middle [of] Tianjun stars, 
[there are] fears [that] fire will strike main silo/s [and] granary/ies.
\end{quotation}
Note that both Lou and Tianjun are mentioned.
Although the latter is given only in the omen, Tianjun is explicitly mentioned
in the observational record of Song shi; and it is clear that such an omen would
not be quoted, if the guest star would not have appeared at this location.
Given that `Tianjun' means `Celestial Granary', the omen is related to silos and granaries.

Lou is either the small (lunar-mansion-)asterism 
with $\alpha,\beta,\gamma$ Ari,
or one of the 28 Lunar Mansion right ascension ranges, here the one defined from the determinative star of Lou ($\beta$ Ari)
to the next one (Wei with 35 Ari), i.e. $\alpha$=1h\,4m\,22s to 1h\,50m\,22s;
Mira with $\alpha$=1h\,32m\,44s lies within this range (all for epoch of date, AD 1070 Dec 25).
While Stephenson \& Green (2009) reject the lunar-mansion-range possibility here,
there are in fact many guest star and comet reports, where the lunar mansion right ascension range is indeed meant
(without any further specification),
when one of those 28 terms like Lou is given 
(e.g., D.L. Neuh\"auser et al. 2021; R. Neuh\"auser, D.L. Neuh\"auser, Chapman 2021).
A simple and fully consistent solution would be that this guest star was located in that part 
of the asterism Tianjun, that is also part of lunar mansion Lou (Fig. 1).
However, since the above text has `in the middle [of] Tianjun stars' (in the omen),
Mira does not fit, because it might be the last southern star of Tianjun (see iv and Fig. 1).

\smallskip

The transmitted texts could point to a comet,
because the sky location
is close to the ecliptic and not far from the Sun around Dec 25. 
The term `ke xing' translated as `guest star' is used in comet reports, 
in particular for the first and/or last detection (often unresolved),
see e.g. R. Neuh\"auser, D.L. Neuh\"auser, Chapman (2021).

Other observations of a comet in AD 1070/71 are extant:
\begin{quotation}
\noindent In Armenian year 519 [AD 1070 March 4 to 1071 March 3] a comet appeared in
the sky. Those who saw it said that it is the sign that was seen before, after which
bloodshed happened
\end{quotation}
(cited after Kronk 1999, p. 523, from an Armenian chronicle compiled AD 1137). 
Kronk listed this observation only in his appendix of uncertain comets, 
apparently because the AD 1070 event was suggested as nova by Hsi (1957). 

In Pingré (1783, p. 378), we find a link to the comet chronicles of Georg Caesius (born 1543 in Rothenburg, studied at U Wittenberg, worked
as pastor, wrote yearly calendars until his death in 1604 in Burgbernheim, all in present-day 
Germany, see Kempkens 2016) in Latin and German in Caesius (1579a and 1579b) --
Caesius wrote as follows:
\begin{quotation}
\noindent Latin: Postea anno 1071. Stella in solita in Austram \& occidentem visa est, per dies 25. 
\& post alia miracula Cometes, longos \& flammeos crines ducens,
apparuit. Iohan. Praetorius. \\
German: Hernach im 1071. Jar / ist gegen mittag und Nidergang / ein ungewonlicher Stern 25. tag gesehen
worden / und andere wunderzeichen auch ein Comet / so lange flammende Haar sich gezogen / erschinen. Praetorius.
\end{quotation}
Here our translation:
\begin{quotation}
\noindent Afterwards, in the year 1071, an unusual star [stella] was seen towards the south and west for 25 days,
after other wonders, [including] a comet [that had] appeared with long fiery hair carrying. Praetorius.
\end{quotation}
Caesius specified Johannes Praetorius (1537-1616, 
professor of mathematics/astronomy at U Wittenberg, Germany, see Folkerts 1996) as his source,
and, indeed, we found almost the same text in his Latin and German comet chronicles 
(Praetorius 1578a,b).\footnote{Latin: `Postea anno 1071. stella insolita, in Austrum \& Occidentem 
visa est, per dies 25. \& post alia miracula Cometes, longos \& flammeos crines ducens, apparuit'
(Pratorius 1578a); 
German: `Hernaach im 1071. Jar / ist gegen Mittag und Nidergang / ein ungew\"ohnlicher Stern /
inn 25. tag / gesehen worden / unnd oder andere wunderwerck / auch ein Comet / so lange
stammende har nach sich gezogen / erschinen' (Praetorius 1578b).}
According to the Latin wording (`post') and the context, the comet `with long fiery hair' appeared
(together with the `other wonders') {\it before} AD 1071, almost certainly 1P/Halley in AD 1066
(see descriptions in Kronk 1999 as `fiery' and `hairy').
Also, the Armenian source might reflect on 1P/Halley AD 1066 
(`Those who saw it said that it is the sign that was seen before').
The description as `stellam in solita' (`unusual star') points in the given context also to a comet sighting.
 
Note that the date given from China, converted to AD 1070 Dec 25 ({\it jul.}),
corresponds to the start of year AD 1071 according to medieval western European convention,
where the year started on Christmas (since AD 800).
The dating of the chronicle of Georg Caesius seems correct around this time.\footnote{Shortly 
before the comet of 1071, he reports a large comet of 1066 (1P/Halley) correctly,
and just after 1071, he gives AD 1080 Oct 15 as date for the battle between Emperor Henry IV and
anti-king Rudolf of Saxonia in Hohenm\"olsen, Germany, which is conventionally dated just 
one day earlier (just a few years earlier, Henry IV undertook the Walk of Canossa 
to reconcile with Pope Gregory VII). Also the dating in the chronicle of Praetorius is correct:
He dated both the comet of 1066 
(Halley) correctly as well as the death of Emperor Henry IV in AD 1106.}

The overlap of the given time spans is then AD 1070 Dec 25 (China) to 1071 Mar 3 (Armenia).
If the object was seen for `25 days' (Caesius and Praetorius) since 1070 Dec 25 (China),
then is was probably mostly visible in January 1071.
The given directions (`south and west') point to motion; 
the dating would point to an evening comet before conjunction with the Sun.

The term given in the Armenian report translated as `comet' does not necessarily point to
a comet in today's sense, but could be used as general term for {\it transient} celestial objects.
The term used in the reports by Praetorius and Caesius, `stella in solita' (`unusual star'),
can also point to either a comet in today's sense or a truly stellar object (today: variable or new star);
the end of their sentences (`long fiery hair') clearly points to a comet (1P/Halley 1066).
Still, the fact that `south and west' are given points to motion and, hence, a comet (instead of a star) for AD 1070/1.
In addition, the visibility period of 25 days is not atypical for comets, but is not expected for Mira, which
either has a bright maximum (ca. 2nd mag) and is then seen for 2-3 months, or, if the maximum was much fainter,
it would not have been discovered serendipitously.
In sum, since the various transmissions from different parts of the world give a consistent picture
for a comet, the guest star of AD 1070 is less likely a sighting of Mira.

\bigskip

{\bf (iv) Was Mira a constituent of the asterism Tianjun?} \\
Finally, we consider whether Mira was part of any of the asterisms in the Classical Chinese sky.
Indeed, according to some reconstructions of the constituent stars, 
Mira could be one of the stars of Tianjun (less likely Chuhao).
Tianjun is established with 13 constituent stars connected with a somewhat curvy, serpentine skeleton line;
the stars in the NW and middle are 3-4 mag ($\alpha,\kappa,\lambda ...$ Ceti)
and are mainly the same stars as in the head of Cetus (see Fig. 1).
Some of the constituents seem to have changed over time, e.g. on the Song Su and temple charts
(Fig. 1 upper panels), EL Ceti may be more likely than $\kappa^{1}$ Ceti.
Also, the connecting skeleton line can partly be different, 
even though of probably the same constituent stars (see Song Su and temple charts). 
We concentrate here on the lower end (SW) of Tianjun.

According to the reconstructions of the Chinese asterisms given in Pan Nai (1989, p. 405),
based on other works mainly from the 17th to early 20th century and partly on measurements from AD 1049-1054,
the asterism Tianjun ends in the south with either 66 or 70 Ceti or {\it o} Ceti (as star no. 13 in modern counting).
Some identifications of the last stars in Tianjun may be problematic,
because they are rather faint (63 Ceti at V=5.9 mag, 
66 Ceti at V=5.7 mag, 
84 Ceti at V=5.8 mag, ESA 1997). 
The reconstruction by Yi Shitong (1984), see Fig. 1 (lower right), is problematic because of rather faint stars,
but also the small separation between 63 and 66 Ceti does not mirror well the maps. 

In the reconstruction by Sun \& Kistemaker (1997), also in Fig. 1 (lower right, full lines),
Mira is part of Tianjun as its southernmost star
after $\delta$, 75, 69, and 70 Ceti -- this combination reflects the Ming chart (Fig. 1, lower left), 
but not the pattern seen on the earlier Song Su and temple charts (Fig. 1, upper row), 
where only $\delta$, 75, 70, and {\it o} Ceti seem to be depicted.  

In addition to the mentioned charts and similar ones, there are also some alternative reconstructions
of Tianjun, where we cannot localize Mira (e.g. unrealistic pattern or uncertain positions).
Note that all such reconstructions are based only on explicit coordinates of one or very few stars
(here not including Mira), and otherwise on ancient maps and textual descriptions. 

Whether Mira is the last star of Tianjun, depends on whether the Chinese astronomers opted for a sometimes bright star 
(Mira, up to ca. 2nd mag and then brightest in its asterism), 
or whether they have accepted a different, but relatively faint star here,
because Mira is strongly variable and more than half of the time invisible.

If the Chinese astronomers considered Mira as part of one of
their constellations, they would not have reported it as a guest star.
There is also no specific record known from East Asia reporting the variability of Mira.
Many Chinese omina mention the variability of stars (specific or in general)\footnote{e.g., 
{\it `When the stars of Kuei are bright it is a sign of a good
harvest of the six grains. Dimness of the stars forebodes a dispersion of population'}
(Ho 1966, p. 102, from the `Jin shu', the history of the Jin dynasty, AD 265-420, written in the 7th century); 
the four stars of Kuei (Pinyin: (Yu-)Gui) are $\delta,\gamma,\eta,\theta$ Cnc.}, even though due to weather,
so that the variable brightness of stars (including their temporal disappearance) 
may not have been a major concern for these scholars.
This was different in the Latin West: According to the later main-stream interpretation of Aristotle's
{\it Meteorology}, all objects beyond the moon (supra-lunar) including in particular the stars should be
eternal and should never change in brightness or number, but see footnote 4 (Pliny on Hipparchos).
According to Jewish and Christian literal reading, the {\it Book of Genesis} would report that all 
stars would have been formed already on the fourth day of {\it creation} (Gen 1, 14-19); 
see, e.g., Fabricius as pastor, who considered regarding the `new star' (Mira) that
`God ... enlightens these invisible bodies, so that they become visible and to come out in public' and
also that they  `later remain on sky, but invisible' (Sects. 3.4 and 3.5 letter (iii), respectively).
For a discussion of these issues in the context of the new star of AD 1572, see Weichenhahn (2004).

\section{The texts: Historical reports by David Fabricius}

We document here the original texts and provide our English translations.
The texts are found in the diary by Fabricius (Sect. 3.1),
in a Latin letter by David Fabricius to Tycho Brahe (3.2),
in letters by Fabricius to Johannes Kepler (3.3), 
in Fabricius' work on the supernova of 1604 (3.4), and
in a later work in Old German by Fabricius from 1614 (3.5).

The relevant data by Fabricius for Mira for 1596 (observational dates, brightness, and color)
are compiled in Table 1 in Sect. 3.6, where we also discuss a few related issues
like calendar reform, observing time, and in particular Fabricius' theory for the appearance of Mira
in the historical context of the first detection (1596) and the re-detection (1609). 

\subsection{Diary by Fabricius (on 1596)}

In his diary, David Fabricius wrote on Mira as follows, first the modern German citation in Bunte (1888), 
then our English translation (round brackets from Fabricius, square brackets from us):
\begin{quotation}
\noindent Als ich am 3. August vet. st. 1596 fr\"uh morgens Jupiter beobachtete, sah ich gegen S\"uden
einen hellen Stern, der etwas gr\"osser war als die drei Sterne am Kopfe des Widders, und zwar von roter Farbe.
Er stand in 25. 47 (s\"udliche Breite $15^{\circ}54^{\prime}$).
Jupiter war damals in der Meridianh\"ohe beim Aufgang der Sonne $50^{\circ}7^{\prime}$.
Am 11. August vet. st. mass ich mit dem Quadranten die Meridianh\"ohe dieses Sternes und fand 31 gr. 30 min.
Jupiter war damals von ihm entfernt $20^{\circ}35^{\prime}$. Er war ein Stern zweiter Gr\"osse. Im Oktober verschwand
er wieder, und bald nachher folgte die allgemeine Pest in Europa.
\end{quotation}
Bunte (1888, pp. 4-5). We translate:
\begin{quotation}
\noindent When I observed Jupiter on 1596 August 3 old style [Julian] in the early morning 
I saw towards south
a bright star, which was slightly larger [brighter] than the three stars on the head of Aries, and of red color.
It stood in [ecliptic longitude Aries] 25th [degree] 47 [minutes] (southern [ecliptic] latitude $15^{\circ}54^{\prime}$).
Jupiter was then at sunrise in the meridian altitude of $50^{\circ}7^{\prime}$ [see footnote 10 below].
On August 11 old style [Julian] I measured the meridian altitude of this star with the quadrant and found $31^{\circ}30^{\prime}$.
Jupiter was then separated from it by $20^{\circ}35^{\prime}$. It was a star of 2nd magnitude.
In October it disappeared, and soon thereafter the general pestilence in Europe followed.
\end{quotation}
What was called {\it Fabricius' diary} (Bunte: `Tagebuch') was in fact a summary of his notes.

When Mira disappeared in October cannot be derived more precisely from the `pestilence in Europe' --
pestilence deaths are known already for Aug/Sep 1596 elsewhere (e.g. \url{https://zeitreise-bb.de/pest});
in his diary, Fabricius has clusters of death reports on 1596 Sep 5, 11, 14 and Nov 12 \& 13 ({\it jul.}),
see Bunte (1885, pp. 107-108); see also Sect. 3.4.

\subsection{Letter by Fabricius to Brahe (on 1596)}

We quote here in full the letter from David Fabricius to Tycho Brahe (1546-1601, Danish astronomer,
who observed the supernova of 1572 and the comet of 1577 intensively), 
as cited by Brahe.
Later, David Fabricius visited Tycho Brahe twice, once in May 1598 in Wandsbek near Hamburg, Germany,
after Brahe had just left Denmark,
and once May/June 1601 in Prague, where Brahe now worked as imperial astronomer (Thoren 1990).
After his own observations in 1597,
Brahe brings two sets of observations from Fabricius, first from 1597, then on Mira from 1596;
then, Brahe continues with his own observations in 1598. 
We recruit on the edition of Brahe's works by Dreyer (1913, pp. 113-116),
round brackets as given there:
\begin{quotation}
\noindent Observationes in ascititia quadam Stella in asterismo
Ceti anno 96 apparente habitae. \\
Anno praedicto die 3 mensis Augusti observaturus matutino tempore Iovem eiusque distantias a
vicinis stellis insignioribus (quia ob aestivum aerem \& auroram minimae vix apparebant) per
instrumentum meum observaturus, conspexi versus meridiem in asterismo Ceti insolitam, \& antea
ea magnitudine in isto loco non visam stellam, cuius aspectus diligens \& loci consideratio
suspitionem de novo Cometa exorto statim mihi movit. 
Inspiciebam mox globum meum stelliferum, perlustrabam canonem stellarum Prutenicum,
an forte eius magnitudinis stella illic existeret, sed nihil reperi, quod ad locum, multo minus
ad magnitudinem visam quadraret. \\
Hora igitur 1~1/2 ante solis ortum die praedicto
distantiam Jovis \& Aldebar. accepi exactam Gr. 24 Min. 9. Distantia Jovis \& stellae clarae
versus meridiem (nam in hunc modum, cum nihil certi de illa mihi constaret,
stellam novam notabam) 20~22$^{\prime}$ circiter. Sole iam exorto altitudinem Meridianam exactam
Jovis per quadrantem reperi Gr. 50 M. 2. \\
9 Augusti mane circa idem tempus distantia Jovis \& Aldeb. erat 23 Gr. 55 minut.
Distantia Jovis \& stellae versus Meridiem vel stellae novae 20 Gr. 31 Min., altitudo
Meridiana Jovis in ipso solis ortu exacta erat Gr. 50 M. 7. \\
11 Augusti mane altitudinem huius novae stellae (a qua tanquam incognita hactenus
Jovem observaveram) accepi exactam Gr. $30^{\circ}$\footnote{Dreyer already noted this typo: `like this
in the manuscript, should be $31^{\circ}$' --
given the measurement on, e.g., Aug 14 (below), the diary (Sect. 3.1), and the true meridian altitude of Mira 
at Resterhafe, where Fabricius observed, which is $31^{\circ}33^{\prime}$
at meridian passage shortly before sunset.} M. 31 fere. \\
Distantia Jovis \& novae stellae tunc erat Gr. 20~35$^{\prime}$, stella haec erat in Meridiano
cum lucida Arietis iam eundem duobus circiter gradibus transiisset. \\
14 Augusti exactam habui observationem mane: Distantia Jovis \& Aldeb.
23~43, altitudo Meridiana Jovis $50^{\circ}~12^{\prime}$. Jupiter \& nova stella distabant Gr. 20 M. 36, altitudo
huius stellae novae $31^{\circ}~31^{\prime}$. Distantia huius stellae a cauda Ceti 27~50$^{\prime}$ fere.
Haec nova stella \& mandibula Ceti 12 Gr. 51$^{\prime}$.  
Eadem nova et lucida Arietis $26^{\circ}~36^{\prime}$ (addidi in scriptis meis observationem hanc:
Haec stella secundae magnitudinis est, paulo maior lucida Arietis, rubens ut Mars). \\
17 Augusti mane distantiam huius novae stellae \& lucidae Arietis accepi 26~37$^{\prime}$.
Novae \& mandibulae $12^{\circ}~50^{\prime}$ an vitio observationis evenerit, ut nunc uno minuto
distantia lucidae mandibulae \& novae stellae discreparet, non scio. Quod facile
in instrumento duas rimulas tantum habenti fieri potuit. Eodem tempore distantia Jovis \&
Aldeb. erat G. 23. 42, altitudo meridiana Jovis 50~12$^{\prime}$ vel paulo plus. \\
21 Augusti mane distantia Jovis \& Aldeb. 23~40$^{\prime}$. Respondebant tunc distantia
lucidae Arietis \& Cometae 26~37, Cometae \& mandibulae 12~51 vel 50~1/2. \\
Differentiam observationum distantiarum novae stellae \& lucidae Arietis \& mandibulae
(quae aliquando erat uno minuto maior vel minor ut ex praepositis
observationibus liquet) credo ex refractionibus (qua me nunc instruxisti) originem habere,
quod videlicet aliquando citius, aliquando paulo ferius easdem distantias observaverim.
Nam mandibula Horizonti propior maiorem refractionem \& proinde minorem distantiam
dabit, \& si forte hoc modo in distantia Cometae \& lucidae differentia excusari non possit,
quod altitudo lucidae circa illud tempus refractionis expers fuit, puto tamen sic excusari,
quod nova stella sublimior facta minorem refractionem habuerit \& perinde
distantiam a lucida maiorem. Sed T. E. pro subtili suo ingenio haec omnia \& singula
diligentius rimabitur. \\
Post 21 Augusti instrumentis locum ipsius non observavi amplius, quia nullum motum ex
praecedentibus observationibus animadvertere potui. Vidi tamen praedictam stellam
postmodum aliquoties usque ad dies priores Septembris, sed quotidie quasi diminui. Ex
recessu Jovis, cuius illuminatione eam magnitudinem habere praesumebam, nunc item
diminui existimabam, \& propterea amplius, ut forte fixam exstantem non curabam, nec
etiam propter continuas fere aëris obscuritates caeteros planetas tunc aliquibus
septimanis observabam, nisi quod Jovis distantiam ab Aldeb. 1 Sept. mane invenirem 23~54$^{\prime}$, 
altitudinem eius Meridianam 50$^{\circ}$~4~1/2$^{\prime}$. \\
Si observationes hae de ascititia stella eiusmodi essent, quae ad confirmationem tuae
sententiae facere possent, gauderem utique. Feci tamen quantum potuit tunc fieri propter
causas in literis adductas. Testor autem nihil imo ne iota quidem aliter me scripsisse quam
observarim, quod et distantiae Jovis ad eandem stellam patefacient
\end{quotation}
(Dreyer 1913, pp. 114-115; previously published in Latin in the 
{\it Vierteljahresschrift der Astronomischen Gesellschaft} 1869, pp. 290-292).

Here our English translation (square brackets and footnotes are our additions):
\begin{quotation}
\noindent Observations made for a certain unknown [lat. {\it ascititia}] star that appeared in the year [15]96 in the constellation Cetus: \\
When I wanted to measure in the previously mentioned year [1596], on August 3 
in the early morning [{\it matutino} for `early light'], Jupiter and the
separation [{\it distantia}] of it to its neighbouring important stars, and when I was about to observe them with 
the aid of my [non-telescopic] instruments
(when they did not appear very small due to the summer air and redness of dawn),
I saw in the southern direction in the constellation Cetus an unusual [{\it insolitam}] star that was never seen in this size
[brightness] at that location, so that from a closer inspection of the location I immediately thought that
it could be a newly appeared comet. I inspected soon my starry globe, checked in detail the Prudentinian Canon of 
stars,\footnote{The `Tabulae Prutenicae', or Prutenic or Prussian Tables or Canon, were published by Erasmus Reinhold in 1551,
which include a list of stars very much like the Almagest (and indeed without Mira as in the Almagest);
we consulted manuscript Astron. 296 of SLUB Dresden from 1551.}
whether there is perhaps the star of that size, but I have not found anything, which would fit that location,
much less of the seen size. \\
Therefore, on the previously mentioned day [Aug 3] 1.5 hours before sunrise, 
I have obtained $24^{\circ}9^{\prime}$ as separation between Jupiter and Aldebaran.
The separation between Jupiter and the bright star [{\it stellae clarae}] (this was the way I listed the new star,
because I did not know anything certain about it), 
which was directed towards south, was about $20^{\circ}22^{\prime}$.
After the sun had already risen I have taken the exact meridian altitude of Jupiter with
a quadrant to be $50^{\circ}2^{\prime}$.\footnote{Note the day-time observation
of Jupiter by Fabricius for Aug 3: At culmination,
Jupiter was at $50^{\circ}$ altitude that day at Fabricius' site (and $\sim 100^{\circ}$ separated from the Sun).
The sky brightness at the location of Jupiter 
was $\sim 8.8$ mag/arcsec$^{2}$ with a limiting visual magnitude of ca. $1.1 \pm 0.4$ mag 
for a very experienced observer (see Crumey 2014 and {\url https://www.cleardarksky.com/others/BenSugerman/star.htm},
input data: location Resterhafe at 0m (sea level), at the time of Jupiter's meridian
passage with altitude $50^{\circ}$ on 1596 Aug 13 {\it greg.} with temperature $50^{\circ}$ Fahrenheit and humidity 
$40\%$, with Snellen ratio 1, experience 10 (highest), and age 32 years for Fabricius)
but Jupiter had $-2.3$ mag
(dimmed by ca. 0.2 mag due to atmospheric extinction at airmass 1.3),
Fabricius could detect Jupiter above the limit ca. 30 minutes after apparent sunrise.} \\
On August 9 in the morning [{\it mane}, roughly civil twilight] at about the same time, 
the separation between Jupiter and Aldebaran was
$23^{\circ}55^{\prime}$. The separation between Jupiter and the star, which was directed towards
south, or actually the new star, was $20^{\circ}31^{\prime}$.
The meridian altitude of Jupiter at the very sunrise was exactly $50^{\circ}7^{\prime}$. \\
On August 11 in the morning [{\it mane}], I have obtained the exact altitude of this new star [{\it novae stellae}]
(about which basically nothing was known to me until the point, when I observed Jupiter)
of $30^{\circ}$ and almost $31^{\prime}$.\footnote{See footnote 8. In his diary, he gave $31^{\circ}30^{\prime}$ (Sect. 3.1).}
The separation between Jupiter and the new star was at that time $20^{\circ}35^{\prime}$,
this star was on the meridian\footnote{This was 31 minutes before apparent sunrise, Sun $7^{\circ}$ below horizon.}, 
when the bright [star] of Aries had past it [the meridian] already by about $2^{\circ}$\footnote{When
$\alpha$ Ari was $2^{\circ}$ past meridian, the Sun was $8.5^{\circ}$ below horizon}. \\
On August 14 in the morning [{\it mane}], I had a precise observation:
The separation between Jupiter and Aldebaran was $23^{\circ}43^{\prime}$, the meridian altitude
of Jupiter $50^{\circ}12^{\prime}$. 
Jupiter and the new star [{\it nova stella}] were separated by $20^{\circ}36^{\prime}$,
the altitude of the new star was $31^{\circ}31^{\prime}$. The separation of this star from the tail
of Cetus [{\it cauda Ceti}, i.e. $\beta$ Ceti] by now was almost $27^{\circ}50^{\prime}$. 
This new star and the lower jaw of Cetus [{\it mandibula}, i.e. $\alpha$ Ceti] [were separated by]
$12^{\circ}51^{\prime}$. 
Namely this new [star] and the bright [star] of Aries [$\alpha$ Ari] [were separated by] $26^{\circ}36^{\prime}$.   
(I have added to my notes then this observation: 
This star is of 2nd magnitude, slightly larger [brighter] than the bright [star] of Aries,
[and] red [{\it rubens}] like Mars.) \\
On August 17 in the morning [{\it mane}], I have obtained as separation between this new star [{\it novae stellae}] and the
bright [star] of Aries $26^{\circ}37^{\prime}$. I do not know whether the separation of
the new [star] and the lower jaw [$\alpha$ Ceti] of $12^{\circ}50^{\prime}$ was a result of an
observational mistake,
so that now the separation of the bright star [$\alpha$ Ari], the lower jaw [$\alpha$ Ceti], 
and the new star [Mira] differ by one minute.
This can happen easily with an instrument, which has only two marks.
At that time the separation between Jupiter and Aldebaran was $23^{\circ}42^{\prime}$,
the meridian altitude of Jupiter was $50^{\circ}12^{\prime}$ or a little less. \\
On August 21 in the morning [{\it mane}], the separation between Jupiter and Aldebaran was $23^{\circ}40^{\prime}$.
The separation between the bright [star] of Aries [$\alpha$ Ari] and the comet [{\it cometae}, i.e. Mira] corresponded to $26^{\circ}37^{\prime}$,
the [separation] of the comet and the lower jaw [$\alpha$ Ceti] $12^{\circ}51^{\prime}$ or $[12^{\circ}]50.5^{\prime}$. \\
I think that the difference of the observations of the separations of the new star and the bright star of Aries
and the jaw (which was sometimes larger or smaller by one minute, as is clearly seen in the previous observations)
has its origin in the refractions, because I have sometimes observed these same separations faster,
and sometimes somewhat more roughly.
Because the jaw, lying closer to the horizon, will show a larger refraction and therefore a
smaller separation, even if perhaps the difference in separations between the comet and the bright star
cannot be excused/explained in this way, because the altitude of the bright star around that moment
was free of refraction -- I nevertheless think, that it 
[the difference] can be excused this way,
because the new star, which was made higher, had a smaller refraction and in the same extend then
a larger separation to the bright star. But all this and the individual [details] will be studied
by Your Excellency with Your perceptive mind. \\
After August 21, I did not measure the position of it [Mira] any more with instruments, because I could not notice any motion with
the previous observations. However, I did see the previously mentioned star soon thereafter several times
until the first days of September, but I saw that it diminished daily. \\
Because of the recession [{\it recessu}] of Jupiter [i.e. retrograde motion, see Sect. 3.6], 
from whose light [{\it illuminatione}] I suppose that it [Mira]
got its size [brightness], I thought that it [Mira] now also got diminished therefore,
and therefore I did not observe the other planets so much for a few weeks, as I 
also did not care much any more about the randomly existing fixed [star, i.e. Mira] -- also because of the
almost continuous darkening of the sky, except that the separation between Jupiter
and Aldebaran on September 1 in the morning [{\it mane}] was $23^{\circ}54^{\prime}$, and a meridian
altitude of $50^{\circ}4.5^{\prime}$.
If these observations of the unknown [{\it ascititia}] star are of some kind, that could help confirming your [Brahe's]
opinion\footnote{Probably that transient celestial objects like new stars and comets
are supra-lunar, new stars even outside the solar system.}, then I would be happy.
I have done what was then possible -- for the reasons given in the letter.
I testify from the depth of my heart that I did not write it differently by an iota than
what I have observed, as the separations between Jupiter and that star show.\footnote{This text
was partly translated to English by Rosen (1967) -- and similar by Hatch (2011, note 10) following Rosen;
however, there might be a problem in one essential sentence: For Latin `... Jovis, cuius illuminatione eam 
magnitudinem habere praesumebam', they give (with omissions before and after this sentence)
`... I sought to judge its [Mira's] magnitude by the brightness of Jupiter ...'.
If this should mean that Fabricius estimated the brightness of Mira (ca. 2nd mag) by comparison to 
Jupiter ($-2.3$ mag), it would be incorrect, because he clearly gave the explicit comparison 
with $\alpha$ Ari (2.02 mag). Regarding the connection of Mira to Jupiter, see Sect. 3.6 below.}
\end{quotation}

When the object was seen first on Aug 3, Fabricius thought that it could be a new comet
(e.g. appearance close to ecliptic and around sunrise).
He called it variously a `bright', `unknown', or `unusual star', or `nova stella' or `cometa[e]'.
Since the object neither moved relative to the stars nor developed a tail,
he doubted his first intuitive (comet) interpretation.
That he still called is `cometa[e]' on Aug 21 may show that he was still uncertain
about the classification; and/or he used `comet' as general term for both what we today
consider comets and new or variable stars, but see his later works in Sect. 3.4 with
`nova stella' for a star that can become visible and invisible (also Sect. 3.3 letter iii). 
The understanding of and a differential terminology for such objects was just developing,
in particular with the new stars of AD 1572 and 1604 (see Hoskin 1977, Weichenhahn 2004).

Fabricius seems to have connected the diminishing of Mira after Aug 21 with the retrograde
motion of Jupiter since that date, see discussion in Sect. 3.6.

\smallskip

Let us consider the dating of the letter(s) from Fabricius to Brahe:
According to Bunte (1885, p. 107), Fabricius wrote in his diary that he had sent a first letter 
to Brahe already on 1596 Aug 11 ({\it jul.}), i.e. on Aug 21 ({\it greg.}): \\
`1596 ... August 11. Scripsi primo in dania ad Tychonem, 17. Jupiter stationarius, 29. Rex daniae coronatus est', 
i.e. `1596 ... August 11. Written [for the] first [time] to Tycho [Brahe] in Denmark,
17th Aug Jupiter stationary, 29th Aug King of Denmark crowned'. \\
Then, he also wrote \\
`1596 ... Sep 29, literas Tych. accepi', \\
i.e. `1596 ... Sep 29, letter received from Tycho', see Bunte (1885, p. 107) for details,
his Sep 29 ({\it jul.}) is Oct 9 ({\it greg.}).
Fabricius received another letter from Brahe in Nov 1596 (Bunte 1888, p. 6).
It is therefore conceivable that Fabricius wrote another letter to Brahe in between:
Indeed, since he gave his last observation of Mira above already for September,
but otherwise (later) for October (Sects. 3.3-3.6), it is well possible that he wrote the above cited letter 
some time in September 1596 ({\it jul.}) --
probably late September, because he mentioned that he did not care much about the new star for `a few weeks' (after Aug 21 {\it jul.});
and he then got a reply from Brahe in November. 
All these original letters are lost except the cited letter from Fabricius by Brahe above.

Brahe did not report to have observed the new star himself:
He observed the bright stars in Cetus in Jan 1582, and later the fainter ones in 1585 Jan, 1586 Jan,
1588 Jan \& Dec, 1590 Nov \& Dec, and 1592 Jan (see Dreyer's edition of Brahe's works).
If Fabricius observed Mira during a maximum in the first half of 1596 Aug ({\it jul.}), 
then Brahe could have detected Mira best in 1582 Jan (when extrapolating backwards with its period of 331 days), 
but not during the later observations in Cetus (Argelander 1869, p. 6). 
In the summer of 1596, Brahe observed the comet of 1596 (in Lyn, UMa, and Leo)
on July 14, 15, 16, 17, 21, 24, and 27 ({\it jul.})
with very precise positions and a drawing in relation to the Big Dipper --
partly from Copenhagen and partly from Hven, both present-day Denmark (Dreyer 1913, vol. XIII, pp. 390-393).
Fabricius reported precise coordinates for this comet already for 
July 10 and observations also for July 8 and 16 ({\it jul.}) in his diary (Bunte 1885).

Brahe had received the first letter from Fabricius on the new star 
probably in late August or early September (reply received Sep 29);
and the second letter not before the end of September.

Brahe's meteorological diary (partly compiled by others) shows that he could hardly observe in
August 1596: Brahe had left Hven for 1596 Aug 4-15 ({\it jul.});
and most employees had left Hven 1596 Aug 28 to Sep 8 (but Brahe returned only on Sep 15) --
probably for the week-long coronation ceremonies of the new King of Denmark, Christian IV, 
in late August 1596, which were attended by Brahe and his entourage, see Thoren (1990, p. 367).
Brahe's meteorological diary records `clear' (day-time) weather for most of the first half of Aug 1596, but a `storm' on Aug 6
and `rain' on Aug 12, 13, 18, 19, 22, 24, 26, and 29 {\it jul.} (Dreyer 1913, vol. 9, pp. 142-143).

For the summer of 1596, Brahe (and colleagues) reported stellar and planetary observations for July 18 and 25 as well as Aug 8,
then again since Oct 1, all {\it jul.} (Dreyer 1913, vol. XIII, pp. 36-37);
plus the comet July 14-27 (see above). Interestingly, for 1596 Aug 8 ({\it jul.}) morning (`mane'), 
Brahe (or, more likely, one of his assistances) measured the separations between Jupiter and the stars $\alpha$ Peg (`prima alae Pegasi'), 
$\alpha$ Tau (`Aldebaran'), and $\alpha$ Ari (`lucida Aries') just before sunrise (`incertum propter Auroram') 
-- Fabricius listed separation measurements between Mira, Jupiter, and Aldebaran as well as Cetus and Aries stars for 
Aug 3, 9, 11, 14, 17, 21 and Sep 1 ({\it jul.}) and compared the brightness of Mira to $\alpha$ Ari.
Note that full moon was on 1596 Aug 8 ({\it jul.}) in Cap,
then decreasing slowly in Aqr (Aug 9 and 10) and on Aug 11 to 13 in the
southern parts of Psc (with a lowest separation to Mira around midnight Aug 11/12 with $18^{\circ}$),
so that observations and precise measurements were more difficult --
but still done by Fabricius. However, Brahe and his assistances apparently did not detect Mira.

\subsection{Letters by Fabricius to Kepler (on 1596 and 1609)}

Next, we quote from letters by David Fabricius to Johannes Kepler (1571-1630, imperial astronomer in Prague
since 1601,
discoverer of the laws of planetary motion -- partly based on planetary positions by Brahe and Fabricius) 
regarding Mira in Latin from the edition of Kepler's work 
by Caspar (1951, 1954 --
slightly modified for better readability, e.g. regarding letters u and v and some abbreviations).

\smallskip

(i) Letter from 1604 Oct 27 ({\it jul.}):
\begin{quotation}
\noindent Nam anno 96. 3 Augusti etiam novam stellam 2 magnitudinis vidi ... quae in Octobri evanuit
\end{quotation}
(Caspar 1951, vol. 15, pp. 58-59; similar in Frisch 1859, vol. 2, p. 597), 
i.e. that the new star seen since 1596 Aug 3 ({\it jul.}) was of 2nd mag and that it disappeared in October.

\smallskip

(ii) Letter from 1605 Jan 14 ({\it jul.}):
\begin{quotation}
\noindent Agnosco sane idem, (testimonium tuum in Opticis de novis stellis) et ut de illa stella anni 96 certi quid habeas,
adjungo observationes Domino Tychoni aliquando communicatas.  
Cum anno 96. 3 Aug. V. St. mane Jovem observarem, vidi claram stellam versus meridiem, paulo majorem stellis tribus in capite Arietis 
eratque rubej coloris ... 11 Augusti stellae huius novae altitudinem meridianam quadrante capiebam $31^{\circ}$ Gr. $30^{\prime}$ M.
... erat 2 magnitudinis. Vide hae observationes certae sunt. Post Michaelis festum disparuit
\end{quotation}
(Caspar 1951, vol. 15, p. 117; similar in Frisch 1859, vol. 2, p. 598, his addition in round brackets).
We translate to English:
\begin{quotation}
\noindent I know your testimony (in {\it Optics} about the new stars) and, to be sure, also what you have about that star in the year [15]96, 
I add the adjunct observations communicated to Mister Tycho.
When I wanted to observe Jupiter on [15]96 August 3 [{\it jul.}] in the morning, 
I saw a bright star towards the south/meridian, slightly larger than the three stars in the
head of Aries, and it was of red color ... 
On Aug 11 I obtained the meridian altitude of this new star as $31^{\circ}30^{\prime}$ ...
It was of 2nd magnitude. These observations are certain. After the feast of Michael, it disappeared.
\end{quotation}

\smallskip

(iii) Letter from 1609 March 12 ({\it jul.}) from Fabricius to Kepler:
\begin{quotation}
\noindent Nova. Corollarij loco. Cum 5 Febr. conjunctionem Jovis et Veneris futuram observarem, animadverti in Caeto stellam insolitam,
quam statim observavi. cum in globo quaererem distantias, vidi eas convenire ad locum stellae in globo annotatae,
quam anno 96 in Aug. et Sept. observavi. quae ab eo tempore a me visa non erat.
res mira. testor Deum me ita bis diversis temporibus vidisse et observasse. et quod notandum, 
Jupiter fere ad eundem locum in Tauro [or: Tauri] pervenerat,
in quo anno 96 erat. non possum satis mirari Dei Opera admiranda et vides hinc mi Keplere, meam de novis stellis et cometis 
sententiam esse veram,
quod non de novo creentur, sed priventur saltem interdum lumine et sic cursus suos nihilominus perficiant. 
quando vero Deo visum fuerit nobis aliquid praeter ordinem significare, accendit illa corpora invisibilia ut appareant
et in publicum tanquam feciales quidam prodeant. cogita tu de his ulterius.
ego puto, me non falso coniectasse antea de istis corporibus aethereis. 
in fine Februarij 
adhuc vidi et observavi certissime. nunc ob minus defecatum caelum et propinquitatem ad horizontem, item 
Lunae radios, animadvertere non potui,
quaeso an vos eam quoque videritis, aut quemquam observasse audiveritis. sententiam
tuam de his scire aveo. res mira et vera. locus ejus (ut in tractatu de nova stella Germanico scripsi)
in $25^{\circ}47^{\prime}$ Arietis. lat. austr. 15$^{\circ}.54^{\prime}$
\end{quotation}
(Caspar 1954, vol. 16, pp. 232-233);
mostly similar in Frisch (1859, vol. 2, pp. 603-604).

Here our literal translation to English (our additions in square brackets):
\begin{quotation}
\noindent New [star]. Instead of a conclusion: 
Since Feb 5, when I observed Jupiter and Venus prior to the upcoming conjunction,
I noticed in Cetus an unusual star, which I observed immediatelly.
When I studied the separations of sky, I saw that they came together at the location of the noticed star,
which I already had marked on the [celestial] globe, and which I had observed in Aug and Sep of the year [15]96, 
and which was not seen anymore by myself since then.
Wonderful thing [lat. res mira]!
I witness to God that I have seen and observed it twice at different times,
and this has to be remarked, that Jupiter has arrived almost at the same location in/of Taurus which it had in the year [15]96.
I cannot wonder enough about the admirable work of God, and You can then see, my Kepler, that my oppinion
on new stars and comets is true, that they are not created anew [lat. ex novo], but they are sometimes deprived of their light,
and nevertheless they follow their course.
But when God finds it advisable to show us something extraordinary, then he enlightens these invisible bodies,
so that they become visible and to come out in public as if they were beauties.
Think about it further.
I think that I have not wrongly considered before regarding those celestial bodies.
To the end of February [probably meaning the last third of Feb, because his
last Mira detection date was Feb 22, see Sect. 3.5], I still saw it, and I observed it certainly.
Now I could not notice it [anymore], not so much due to an imperfect sky and the close horizon,
but because of the rays of the moon [probably the Sun, see below and Sect. 3.5].
I ask, whether You also have seen or heard about it, that someone has observed it.
I desire to know Your opinion on those [matters]. A wonderful [lat. mira] and true thing! 
Its location, (as I have written in a German treatise on the new star,) 
is at $25^{\circ}47^{\prime}$ of Aries and at 15$^{\circ}54^{\prime}$ southern latitude.
\end{quotation}
Fabricius here called the story about the new star twice `res mira' (see footnote 1).
Regarding the connection to Jupiter, see Sect. 3.6.
The given Taurus is the correct ecliptic longitude range for Jupiter at this time.

In the paragraph just before the one cited above, on other matters, Fabricius mentioned that the 
`overcast sky, windy, and cloudy' (lat. `turbidus aer, ventosus et nubilosus') on 1609 Feb 17, 18, 19, 20, and 21 ({\it jul.}),
which includes the dates with observations of Mira Feb 19 and 20,
and that there was `cold air' (lat. `gelida aeris', probably clear sky) on 1609 Mar 4, 5, 6, and 7 ({\it jul.}).

When Fabricius wrote that he did not detect Mira `because of the rays of the moon', he could mean
ca. Feb 25-28 ({\it jul.}), when the moon is close to Mira.
However, since he wrote the letter on Mar 12 ({\it jul.}) and wrote `Now I could not notice it',
the term `now' can hardly pertain to Feb 25-28.
Although the weather was good (cold) Mar 4-7, and even though Mira was still above the horizon,
Mira was indeed very close to the Sun those days, so that he most likely meant
`because of the rays of the Sun'.
The term transcribed in the letter as `moon' is given only as abbreviated astronomical symbol
for the moon, which might have been mistaken for the symbol for the Sun.
Indeed, in his Prognosticon for 1615 (Sect. 3.5), he wrote that he lost Mira 
when it `was covered by the rays of the Sun'.\footnote{Hatch (2011, endnote 16) and Rosen (1967) translated 
one essential part as follows: `From the end of February I saw and observed it clearly; 
now, because of the hazy sky, the proximity to the horizon, and the moonlight, I have not
been able to observe it'. Here, `from the end of February' for `in fine Februarij' cannot be correct, 
because Fabricius observed Mira last on 1609 Feb 22 {\it jul.} (Sect. 3.5) -- it is `to the end of February'.
Also, the non-detection was not due to `hazy sky (and) the proximity to the horizon' --
Rosen and Hatch did not translate the word `minus' for `not so much';
furthermore, Fabricius mentioned that he had bad weather 1609 Feb 17-21 ({\it jul.}),
but not at the end of February.}

\subsection{Fabricius (1605) on Mira and P Cygni in his work on the supernova of 1604}

In Fabricius (1605), which is mainly on the supernova of 1604 (SN 1604), we also found one paragraph on the
new stars of 1596 (Mira) and 1602 (P Cygni). 
We cite from chapter 3 (p. 24) of the manuscript at the Nieders\"achsische Staats- und 
Universit\"atsbibliothek G\"ottingen\footnote{\url{https://gdz.sub.uni-goettingen.de/id/PPN585888809}},
round brackets from Fabricius, square brackets from us:
\begin{quotation}
\noindent Was ich auch selbst in solchen Sachen etliche Jahr her angemercket / wil ich auch herzu setzen:
Anno 1596. den 3. Augusti, habe ich des Morgens inter observandum, einen newen Stern tertiae
magnitudinis, im 25. gr. 47. min. Arietis, cum latitudine Australi 15. gr. 54. min. wahrgenommen /
und fleissig etlich Tage her notiret, der bi\ss~im Octobr. gestanden / und hernach nicht mehr ist
gesehen worden. Darauff sich al\ss balde denselben Herbst / die Universal langweilige Pest in
Teutschlandt angefangen / und bey nahe ganz Europam durchzogen / Dieweil Germania dem
Himlischen Arieti, nach der Astronomorum meynung / unterworffen seyn soll. Zu welcher Zeit
auch die einnemung Erla, und die grosse Feldschlacht zwischen dem Christlichen Krieg\ss volcke /
und dem T\"urckischen Tyrannen in Ungern sich begeben. Wie dann solche beyde effectus durch
die bleichrote Farbe Saturni \& Martis in diesem Stern zuverstehen gegeben sind. \\
Imgleichen ist von mir / auff anzeigung des Edlen und Hochgelahrren
Herrn Francisici Tengnagel / jetze R\"om. Kays. Mayest. appellation Raths /
und seligen Herrn Tychonis Brahe Tochtermannes (der da\ss mahl in
Frie\ss landt mich besuchet hat) Anno 1602. im Martio, ein newer kleiner
Stern / circa pectus Cygni, im 16. gr. 19. min. Aquarij, cum latitudine
boreali 55. gr. 27. min. observiret worden / welcher auch nochmals zusehen ist.
Dieser Stern / weil es exacte regionis nostrae Horizontem stringeret, unnd
corpore suo (quo ad visum) gleich anr\"uhret / wenn er ins Norden untergehen
sollen / hat neben anderen Sachen und Ortern / auch ohne zweiggel die
Ostfrisische erfolgte unruhe rubicundo \& Martiali suo colore evidenter
angedeutet / wie man solches genugsam erfahren hat ...
Au\ss diesen ist nun vern\"unfftiglich anzunehmen / da\ss auch dieser
jetzige newe Wunderstern [SN 1604] grosse Bedeutung habe /
dieweil er viel gr\"osser / als die zween vorgemelte / und sonderlich /
da\ss er circa locum conjunctionis magne angez\"undet worden.
\end{quotation}
We translate this to English literally (round brackets from Fabricius, square brackets from us):
\begin{quotation}
\noindent As I had noted on these matters for several years, I will also mention here:
In the year 1596, on August 3 [{\it jul.}], during the morning observations, I have noticed a new star
of third magnitude, at [ecliptic longitude] Aries $25^{\circ}47^{\prime}$ with [ecliptic latitude] south $15^{\circ}54^{\prime}$,
and recorded diligently for quite a few days, which stood until into October, and which was not seen afterwards.
Then, soon in the same autumn, the universal long-lasting pestilence started in Germany and pulled through all of Europe.
Meanwhile, Germany was subdued by the heavenly Aries, after the astronomers opinion. 
At this time, also the conquest of Erla and the great battle between the Christian war army and the Turkish tyrant
happened in Hungary. As if these two effects were to be understood by the pale red color of Saturn and Mars in this star. \\
As well was observed by myself, after indication from the noble and highly educated
Mr. Franziscus Tengnagel, now 
appeal advisor to the Roman Emperor [Rudolph II, 1522-1612, reigned in Prague as
Holy Roman Emperor and King of Germany, Bohemia, Hungary, and Croatia] and husband of a daughter
of the blessed Mr. Tycho Brahe (who has visited me then in Frisia), in the year 1602 in March
a new small star, in the chest of Cygnus, in [ecliptic longitude] Aqr 16$^{\circ}$ 19$^{\prime}$
with northern latitude 55$^{\circ}$ 27$^{\prime}$, which is still visible.
This star, which touches exactly our horizon, and his body (which is to view) similarly touches,
when it is to set in the north, has next to other things and places, also without doubt the
East Frisian unrest, evidently indicated a reddish [lat. rubicundo] and Mars-like color [Martiali suo colore],
as was experienced sufficently ...
From this it is to assume reasonably, that also this current new wonderous star [SN 1604]
has a very important meaning, because it is larger than the two previous ones and particular,
that it appeared at about the location of the large conjunction.
\end{quotation}

While Fabricius otherwise several times gave `2nd mag' for Mira (and that it was slightly brighter than $\alpha$ Ari, see Table 1),
it may be surprising that he wrote here that Mira is `a new star of third magnitude' --  
he may have already thought of the new star of 1602 (P Cygni), which he reported next as a `small star';
such a phrase (`small' means faint) was never used for Mira, so that P Cygni was probably fainter than Mira.
P Cygni and Mira are also specified to be fainter than the new star of 1604 [SN 1604].
P Cygni is reported as reddish and of Mars-like color (Fabricius in Latin: rubicundo \& Martiali suo colore);
such a color can then be quantified as color index B$-$V=$1.43 \pm 0.13$ mag, just like Mars,
i.e. the same range, value, and uncertainty as Mars also for P Cygni.

P Cygni was of 3rd mag according to Bayer's Uranometria (1603) as well as Henisch and Brengger in letters to Martin Walser,
which were handed to Kepler (Frisch 1859, vol. 2, pp. 756-767), see Granada (2021) for more details.
Such specifications of `3rd mag' mean that the star was as bright as stars given with 3rd mag in the Almagest.
The 
174 stars with plain `3rd mag' in the Almagest today have a mean V=3.17 mag (standard deviation 0.67 mag)
according to Schaefer (2013);
however, according to Hearnshaw (1999), a 3rd mag in the Almagest corresponds to $3.082 \pm 0.054$ mag today.
The five Cygnus stars with plain 3rd in Almagest have a mean of $2.76 \pm 0.43$ mag.
Hence, the brightness of P Cygni was probably in the range of $\sim$2.3-3.2 mag --
and, thus, bright enough for detection of color.
The brightening of the Luminous Blue Variable P Cygni 
(now V=4.79 mag and B$-$V=0.37 mag, ESA 1997) was probably due to a strong wind outburst 
and dust expulsion, hence, extra reddening (see de Groot 1988).

The final disappearance of Mira is connected here again with the new wave of pestilence (like in Sect. 3.1, hardly datable precisely) as well as
the conquest of castle Erlau, Hungary, by Sultan Mehmed III, which happened on 1596 Oct 12 ({\it greg.}), and the great Battle 
of Keresztes, Hungary, Oct 23-26 ({\it greg.}), where the Ottomans defeated an Habsburgian Christian army (which cleared the path to Vienna).

That Fabricius gave both Saturn and Mars here as comparison objects for the color of Mira, might have been motivated by certain interest,
namely that two negative (bloody) events (`two effects') were narrated in connection to Mira (conquest and battle). 
It is still credible that the true color of Mira is meant: `the pale red color of Saturn and Mars in this star'.
Since otherwise Mira is just compared to Mars (see Table 1), the wording here tends to signal a redness closer
to the lower  
B$-$V color index range of Mars. We discuss the color index in detail in Sect. 3.6.

\subsection{Prognosticon by Fabricius for 1615 (on Mira in 1596 and 1609)} 

Finally, we quote from the mostly German (and partly Latin) {\it Prognosticon} by Fabricius (1614) for the year 1615
(the dedication is dated 1614 Jun 1), where he also reported about Mira -- his last known record on Mira.
\begin{quotation}
\noindent Zu dem ist man zu unserer zeit eigentlich in erfahrung kommen / da\ss~neben den Cometen
unterweilen auch novae stellae non caudatae sich sehen lassen / unnd hernach sich widerumb
verlieren / wie Anno 1572. und 1604. von vilen Astronomis observirt worden.
Ich habe auch di\ss~darneben angemerckt / da\ss~solcher novarum stellarum corpora auch hernach
vere, doch unsichtbar am Himel verbleiben / wie ich solches an einem besondern newen Stern
(der meines wissens von keinen andern Astronomis wargenommen worden) observirt habe /
welcher An. 1596 den 3. Aug. und etlichen folgenden tagen / ut stella secundae magnitudinis et
rubicundi coloris ut Mars sich hat sehen lassen / unnd von mir zu etlich malen observirt worden / 
und seinen locum gehabt im 25. grad / 47. minut. Arietis cum latitudine australi 15. grad, 45. minut. \\
Eben denselben Stern / nach dem er in die zwelff Jahr lang verschwunden oder unsichtbar
gewest / hab ich andermals widerumb zu Gesicht bekommen / Anno 1609. den 5. 12. 19. 20. und
22. Febr. vet. styl. de\ss~abends / und gleiche distantias a vicinis fixis gehabt / als f\"ur 12. Jahren /
ist aber bald hernach widerumb verloschen / unnd mit der Sonnenstralen bedeckt worden
\end{quotation}
(Fabricius 1614, p. 3; cited before in Sch\"onfeld 1883). 

We translate as follows:
\begin{quotation}
\noindent Also, in our time, it became really known that, in addition to comets, also new stars [novae stellae] without tails [non caudatae] 
are seen, which later get lost, like in the years 1572 and 1604, observed by many astronomers.
I have remarked, that the bodies of such new stars also later remain on sky, but invisible, as I have observed
for one such new star (which was not noticed by any other astronomer before, as far as I know),
which was seen in the year 1596 on Aug 3 and several days thereafter, like a star of 2nd magnitude and of
red [{\it rubicundi}] color like Mars, and observed by myself several times, and it had its location in the 25th degree and
47th minute of Aries [ecliptic longitude] with a southern [ecliptic] latitude 15th degree and 45th minute. \\
And this very same star, after it had disappeared for 12 years or was invisible, I have seen it again in the year
1609 on Feb 5, 12, 19, 20, and 22 old style [{\it vet. styl.} for the old Julian calendar,
i.e. 1609 Feb 15, 22, Mar 1, 2, and 4 ({\it greg.})] in the evening, 
and with the same separation to its nearby fixed stars,
as 12 years earlier, and it again disappeared soon thereafter, and was covered by the rays of the Sun.
\end{quotation}
In the last lines, it is not meant that Mira `disappeared' due to, e.g., faintening, {\it and then}
`was covered by the rays of the Sun', but that it was not detectable anymore because of the nearby Sun
(see also Sect. 3.3 iii).

We note that the observations in 1609 were also done with the unaided eye, because David Fabricius
could start with telescopic observations only after his son Johann had brought a telescope from Leiden, The Netherlands,
in 1611, where he had studied.
Since Hans Lippershey, the inventor of the telescope, had submitted his patent request in September 1608,
it is quite remarkable that, already in 1611, Johann Fabricius could bring a telescope from Leiden.

\subsection{Summary of observational details}

We summarize the reported data on Mira for AD 1596 in Table 1.
These basic data are then to be analyzed more quantitatively in Sect. 4.
We add some remarks on the observing times and the dating (calendar reform),
brightness, color, and disappearance of Mira.
We also consider briefly the re-observations of Mira in 1609.
We end with discussing
Fabricius' theory on the brightening 
and re-appearance of Mira.

\smallskip

{\bf Calendar:} As a protestant (like Brahe), Fabricius gave dates on the old Julian calendar, as he mentioned explicitly in
his diary (Sect. 3.1); his 1596 Aug 3 to 21 ({\it jul.}) correspond to Aug 13 to 31 ({\it greg.}).
The Gregorian calendar reform (a jump from 1582 Oct 4 to 15) was accepted in most protestant parts of Germany
and Brahe's Denmark only in AD 1700 by jumping from Feb 18 to Mar 1, and in Frisia slightly later
by jumping from 1700 Dec 31 to 1701 Jan 12 (Grotefend 2007, p. 27).

\smallskip

{\bf Observing times:} The observations by Fabricius on 1596 Aug 3 ({\it jul.}) were specified by him as `matutino' and for
`1.5h before sunrise' (Sect. 3.2), i.e. when the Sun is $11.5^{\circ}$ below horizon (Frisia) 
near the border from astronomical to nautical twilight.
For the other dates, he gave `mane' (`morning', civil twilight), which apparently also included the time shortly
after sunrise, as e.g. the day-time observation of Jupiter on Aug 3.
For Aug 9, he reported for the time period of `mane', that `the bright [star] of Aries had past it 
[the meridian] already by about $2^{\circ}$': when $\alpha$ Ari was $2^{\circ}$ past meridian, 
the Sun was $8.5^{\circ}$ below horizon. 
For Aug 11, he also reported for `mane', that Mira `was on the meridian', and
this was 31 minutes before apparent sunrise with the Sun $7^{\circ}$ below horizon.
Hence, we can assume for the beginning of `mane' roughly the time of late nautical to civil twilight.

\begin{table*}
\caption{{\bf Data on Mira from David Fabricius for 1596:}}
\begin{tabular}{llll} \hline
Property          & Text                    & \hspace{-0.8cm} Section & Value  \\ \hline
Dates             & 1596 Aug 3, 9, 11, 14, 17, 21, Sep 1, 29 (1) ({\it jul.}) & 3.2 & Aug 13, 19, 21, 24, 27, 31, \\ 
                  & (these dates are given explicitly) & & Sep 11, Oct 9 ({\it greg.}) \\ \hline
Brightness        & `slightly larger than the three stars on the head of Aries ... & & \\
at peak           & ... a star of 2nd magnitude'  (2)  & 3.1 & V=$1.9 \pm 0.1$ mag (3) \\
                  & `this star is of 2nd magnitude ... (2) & & \\
                  & ... a little larger than the bright of Aries' & 3.2 & V=$1.9 \pm 0.1$ mag (3) \\
                  & `slightly larger than the 3 stars in the head of Aries ... & & \\
                  & It was of 2nd magnitude' (2) & 3.3 & V=$1.9 \pm 0.1$ mag (3) \\
                  & `a new star of 3rd magnitude' [probably a scribal error] & 3.4 & \\
                  & `a star of 2nd magnitude' (2) & 3.5 & \\ \hline
Color (index)     & `of red color'            & \hspace{-0.8cm} 3.1, 3.3 & B$-$V$\ge 0.8$ mag (4) \\
                  & `red like Mars'           & 3.2 & B$-$V=$1.43 \pm 0.13$ mag \\
                  & `pale red color of Saturn and Mars in this star' & 3.4 & \\
                  & `red color like Mars'     & 3.5 & B$-$V=$1.43 \pm 0.13$ mag \\ 
                  & \multicolumn{3}{l}{not compared in color to nearby Hamal (B$-$V=1.16 mag) and Aldebaran (B$-$V=1.48 mag)} \\ 
                  & & \hspace{-0.6cm} in sum & B$-$V$\simeq$1.3-1.4 mag \\ \hline
Diminishing       & `After Aug 21 [{\it jul.}] ... diminished daily' (5) & 3.2 & after Aug 31 ({\it greg.}) \\ \hline
Disappearance     & `In October it disappeared' & 3.1 & Oct 11-Nov 10 ({\it greg.}) \\
(Julian           & `for a few weeks, as I also did not care much any more & & \\ 
calendar)         & about the randomly existing fixed [star]' (5) & 3.2 & \\
                  & `it disappeared in October' ({\it jul.}) & 3.3 & Oct 11-Nov 10 ({\it greg.}) \\
                  & `After the feast of Michael, it disappeared' (Sep 29 {\it jul.}) & 3.3 & after Oct 9 ({\it greg.}) \\ 
                  & `stood until into October' & 3.4 & Oct 11-Nov 10 ({\it greg.}) \\
                  & `conquest of Erla' and `Battle' of Keresztes, Hungary & 3.4 & 1596 Oct 12-26 ({\it greg.}) \\ \hline
\end{tabular}
\vspace{.1cm}

Remarks: (1) probably also on Sep 29 (St. Michael) as indirectly specified (see 3rd-to-last line), \\
(2) 2nd mag on the Almagest scale, \\
(3) $\alpha$ Ari has V=2.02 mag, \\
(4) Neuh\"auser et al. (2022), table 1. \\
(5) written in September 1596 ({\it jul.}).
\end{table*}

\smallskip

{\bf Brightness of Mira:}
When Fabricius specified for Aug 1596 that Mira was `slightly larger', i.e. brighter, than the three stars on the head of Aries,
i.e. $\alpha,\beta,\gamma$ Ari and `of 2nd magnitude', he stands in the tradition of the Almagest and the Prudentian table
mentioned by Fabricius: $\alpha$ Ari brighter than typical 3rd, $\beta$ Ari 3rd, and $\gamma$ Ari fainter
than typical 3rd (see Toomer 1984 for Ptolemy's Almagest); it is considered that Ptolemy meant one third mag brighter or fainter 
with `brighter than' and `fainter than', respecitvely. 
The modern data of these three stars are: 
$\alpha$ Ari V=2.02 mag, $\beta$ Ari V=2.66 mag, $\gamma$ Ari V=3.88 mag (ESA 1997).
Since the new star is specified to be brighter than those Almagest 3rd mag stars, it is indeed a star of 2nd magnitude.

More precisely, Fabricius reported for Aug 1596 that `this star is of 2nd magnitude, 
slightly larger [brighter] than the bright [star] of Aries' (Sect. 3.2); 
$\alpha$ Ari (Hamal) has V=$2.020 \pm 0.004$ mag (Hipparcos-Tycho star catalog, ESA 1997, 
not variable on GCVS or AAVSO) -- at a separation of $26.6^{\circ}$.\footnote{Hamal has a color index B$-$V=$1.160 \pm 0.004$ mag
according to the Hipparcos-Tycho satellite (ESA 1997), i.e. similar to Mira (see below), so that there should be 
no large offset in magnitude by such a comparison due to the Purkinje and other color effects.}
We conclude that Mira had a (maximum) brightness around $1.9 \pm 0.1$ mag in Aug 1596.\footnote{In the AAVSO
data table for Mira, the data point from 1596 from Fabricius is given with 2.80 mag (without uncertainty,
with reference to Guthnick 1901), which is too faint
given Fabricius' comparison with Hamal; 2.80 mag might have been derived from Ptolemy's Almagest, where Hamal is listed 
with `bright(er than typical) 3rd mag' (today considered to be ca. one third mag brighter than typical 3rd mag) -- 
but Fabricius compared the new star {\it directly} with Hamal 
(`slightly larger than') and gave `2nd magnitude' several times explicitly.
For 1609, AAVSO gives 3.5 mag (without uncertainty,
with reference to Guthnick 1901), which cannot be derived from Fabricius' texts.
Kepler wrote in his `de Stella Nova in pede Serpentarii' (cited here from Frisch 1859, p. 693): `Prima quod David Fabricius,  
quem in observationibus supra quoque fide digmmi celebravi, animadvertit anno 1596 3/13 Augusti (circa quem diem ...) matutino  
tempore novam stellam tertiae magnitudinis', i.e. that Mira would have had 3rd magnitude --
but this is clearly against the information from Fabricius to Kepler (letters from 1604 Oct 27 and 1605 Jan 4, Sect. 3.3).
Fabricius gave `3rd mag' once, but that must have been a mistake (Sect. 3.4).}

Since Fabricius mentioned the faintening of Mira only for after Aug 21 ({\it jul.}), he had not noticed any changes
in brightness from Aug 3 to 21 ({\it jul.}), which is not unusual for the broad maximum (see Fig. 2).
The precision in his brightness specification is better than about one third of a magnitude
(a third of a mag is the precision in the Almagest, see above),
because he gave the maximum brightness as `slightly larger' than $\alpha$ Ari.

\smallskip

{\bf Color:} In all texts for 1596, Fabricius stressed the reddish color of the star, 
namely `red', `like Mars', and `pale red color of Saturn and Mars'.
Fabricius compared the new star in color mostly to Mars, 
even though it was not observable at that time (Aug/Sep 1596) --
very experienced naked-eye observers can have a particularly good color memory.

Interestingly, even though Fabricius compared Mira and $\alpha$ Ari in brightness (`slightly larger'), he did not compare
them in color -- obviously, because Mira was redder than Hamal (B$-$V=1.16 mag).
Although angular separations were measured from Mira to $\alpha$ and $\beta$ Ceti (plus $\alpha$ Ari), 
as well as from Jupiter to $\alpha$ Tau, no color comparison to any of these stars (or Jupiter) was made.

We can quantify these facts such that Mira was reddish like Mars (B$-$V=$1.43 \pm 0.13$ mag):
also, it was not as red as $\alpha$ Tau (B$-$V=1.48 mag), i.e. more near the lower range of Mars,
but redder than 
$\alpha$ Ari (B$-$V=1.16 mag) and $\beta$ Ceti (B$-$V=1.03 mag);
$\alpha$ Ceti is too red (B$-$V=1.63 mag).
Thus, a color index of Mira during maximum of B$-$V$\simeq$1.3-1.4 mag is reliable.
This is also then consistent with Fabricius' specification to see `the pale red color of Saturn and Mars' in Mira
(Saturn has B$-$V=$1.09 \pm 0.16$ mag).

The precision in Fabricius' color specification is then 
as low as or below a few tenth of mag.
Neuh\"auser et al. (2022) came to the conclusion that experienced naked-eye observers like Tycho
Brahe can specify the color index of a star with a precision and accuracy down to $\pm 0.1$ to 0.2 mag.
Murdin (1981) found that naked-eye star color observations by the experienced naked-eye observers M. Minnaert and 
colleagues were consistent with the true B$-$V color indices with an rms scatter of $\pm 0.24$ mag.

\smallskip

{\bf Disappearance:} Fabricius specified that Mira diminished after 1596 Aug 21 ({\it jul.}) and further in September,
and that it disappeared after the feast of Michael (Sep 29 {\it jul.}) in October.
He mentioned the conquest of Erla and battle of Keresztes, Hungary (Oct 2-16 {\it jul.})
as `effectus' (Sect. 3.4) of the color of Mira -- its disappearance might have been shortly before or during these events.

\smallskip

{\bf Re-detection of Mira in AD 1609:}
No explicit information is narrated for brightness and color for 1609 --
just the observing dates are given: 1609 Feb 15, 22, Mar 1, 2, and 4 ({\it greg.}).
In Sect. 4.2, we estimate the sky brightness for Fabricius' observations in 1609
and the limiting magnitudes.

\smallskip

{\bf Excursus on Mira theory:} 
Fabricius speculated that Mira's brightening was related to Jupiter 
(that Mira got its `light' somehow from Jupiter).
Fabricius reported that Mira diminished after 1596 Aug 21 ({\it jul.}).
Jupiter reached stationarity on Aug 21/22 ({\it jul.}), first dated by Fabricius for Aug 17 ({\it jul.})
in his diary; Fabricius' data (Table 2) show retrograde motion since his next measurement on Sep 1 ({\it jul.}).
Fabricius in his letter to Brahe: `Because of the recession of Jupiter, 
from whose light I suppose that [Mira] got its size,
I thought that [Mira] now also got diminished therefore' (Sect. 3.2).
Fabricius thought to have found the reason of Mira's faintening after Aug 21 ({\it jul.}) explicitly in the retrograde motion of Jupiter
and not in Jupiter's brightness evolution;
Jupiter continued to brighten discernably from ca. $-2.5$ mag to ca. $-2.9$ mag by about 1596 Oct 20 ({\it jul.}).
The observational coincidence in 1596 (of Mira's diminishing with Jupiter's regress) led Fabricius to
his Mira brightness 
theory.\footnote{In medieval and early modern times, 
there was a theory that great conjunctions of planets can cause new phenomena to appear
(as apparently happened with the new star of 1604 near a planetary conjunction), 
but that was not in the mind of Fabricius here: 
He thought that Mira is dimishing when Jupiter is retrograde and invisible when Jupiter is absent or far away.
He denied the option `created ex novo' (Sect. 3.3, letter iii).
During the Mira appearance in 1596, there was no conjunction of two or more planets,
but just Jupiter and Mira.
In 1609, there was a Jupiter-Venus conjunction towards the end of March, 
but Mira was then not visible anymore to Fabricius --
and Fabricius considered again Jupiter alone as cause for Mira (Sect. 3.3 letter iii and Sect. 3.4).}
Obviously, this bias constrained the possibility for re-detections:
Fabricius combined the appearance of Mira with a nearby Jupiter in prograde motion.
In a letter to Kepler on his observations of 1609, he wrote  
`Jupiter has arrived almost at the same location in Taurus which it had in the year [15]96'
(Sect. 3.3 letter iii);
Fabricius re-detected Mira in Feb 1609 and, indeed, Jupiter was prograde since 1608 Dec 30.

According to Bunte (1885, 1888), Fabricius was `restless' regarding Mira (Bunte: `ruhelos') sometime before the re-detection in 1609,
which may point in particular to the time since July 1608.
We could speculate that the statement  
might be related to the situation on sky during July 1608: 
Jupiter was in prograde motion until Sep 4 and at about the same ecliptic longitude as on 1596 Aug 3 ({\it jul.}),
but Mira was not yet detected by Fabricius, even though its position was well visible since early July 1608.
Later, Fabricius might have been even more `restless' knowing that Jupiter's next prograde motion since 
end of Dec 1608 would soon conflict with Mira's upcoming conjunction with the Sun.
On 1609 Feb 5 ({\it jul.}), Fabricius did re-detect Mira with Jupiter indeed at about the
same ecliptic longitude as on 1596 Aug 3 ({\it jul.}).
Due to his bias (nearby prograde Jupiter), Fabricius needs one Jupiter period of 12 years for the re-detection
of Mira -- and some luck that this variable star was then indeed seen.

N.B.: During his observing window from 1596 Aug 3 to 14 ({\it jul.}), 
the separation between Jupiter and Mira as measured by Fabricius indeed increased
(and his separation values between Jupiter and Aldebaran decreased until Aug 21
and increased to Sep 1 after stationarity in between, all Julian).

\section{Analysis of Fabricius' observations}

We will now analyse the statements and data by Fabricius regarding positional accuracy (Sect. 4.1),
brightness and period (4.2), color (4.3), and his detection limit for the very red Mira (4.4).
In 1596, Fabricius was working as pastor in Resterhafe (Frisia) 
at $53^{\circ}38^{\prime}$ north and $7^{\circ}26^{\prime}$ east.

\subsection{Positional accuracy}

We can use the separation measurements between Mira and Jupiter or other stars as well
as the meridian altitudes listed by Fabricius to determine his positional accuracy.
We list in Table 2 the data from Fabricius as recorded in Sect. 3 together 
with the true values and the offsets. 
For better comparison with his texts from Sect. 3.2, we use the Julian calendar here.

For Aug 3, we calculated the separations for the Sun 
being $11.5^{\circ}$ below horizon (Fabricius: `1.5h before sunrise', `matutino'),
while for the other dates, we calculated them for the Sun being $8.5^{\circ}$ below horizon (`mane'). 
These are the observing times given by Fabricius (see Sect. 3.6).
Fabricius gave the separation as precise as to the full arc min (and in two cases to half an arc min).
We give the true separations and the offset of Fabricius' values  
with one digit after the comma (Table 2).

\begin{table}[h!]
\caption{{\bf Angular measurements by Fabricius:} We compare various separation 
and meridian altitude measurements by Fabricius in Aug/Sep 1596 with the true values;
offsets are true values minus Fabricius' values.}
\begin{tabular}{lllcc} \hline
\multicolumn{5}{l}{Separation measurements (from Sect. 3.2):} \\ 
\multicolumn{2}{l}{Date ({\it jul.})} &  \multicolumn{2}{l}{\hspace{-.5cm} separation}   & off- \\ 
1596        & between                 & Fabr.               & true       & set \\ \hline
Aug\,3      & Jup\,\&\,$\alpha$\,Tau  & $24^{\circ}09^{\prime}$ & $24^{\circ}11.0^{\prime}$ & $+2.0^{\prime}$ \\
            & Jup\,\&\,Mira           & $20^{\circ}22^{\prime}$ & $20^{\circ}23.5^{\prime}$ & $+1.5^{\prime}$ \\
Aug\,9      & Jup\,\&\,$\alpha$\,Tau  & $23^{\circ}55^{\prime}$ & $23^{\circ}54.4^{\prime}$ & $-0.6^{\prime}$ \\
            & Jup\,\&\,Mira           & $20^{\circ}31^{\prime}$ & $20^{\circ}34.1^{\prime}$ & $+3.1^{\prime}$ \\
Aug\,11     & Jup\,\&\,Mira           & $20^{\circ}35^{\prime}$ & $20^{\circ}40.8^{\prime}$ & $+5.8^{\prime}$ \\
Aug\,14     & Jup\,\&\,$\alpha$\,Tau  & $23^{\circ}43^{\prime}$ & $23^{\circ}45.8^{\prime}$ & $+2.8^{\prime}$ \\ 
            & Jup\,\&\,Mira           & $20^{\circ}36^{\prime}$ & $20^{\circ}39.3^{\prime}$ & $+3.3^{\prime}$ \\
            & $\beta$\,Cet\,\&\,Mira  & $27^{\circ}50^{\prime}$ & $27^{\circ}51.0^{\prime}$ & $+1.0^{\prime}$ \\
            & $\alpha$\,Ari\,\&\,Mira & $26^{\circ}36^{\prime}$ & $26^{\circ}36.3^{\prime}$ & $+0.3^{\prime}$ \\ 
            & $\alpha$\,Cet\,\&\,Mira & $12^{\circ}51^{\prime}$ & $12^{\circ}50.7^{\prime}$ & $-0.3^{\prime}$ \\
Aug\,17     & $\alpha$\,Ari\,\&\,Mira & $26^{\circ}37^{\prime}$ & $26^{\circ}36.3^{\prime}$ & $-0.7^{\prime}$ \\
            & $\alpha$\,Cet\,\&\,Mira & $12^{\circ}50^{\prime \,(1)}$ & $12^{\circ}50.7^{\prime}$ & $+0.7^{\prime}$ \\ 
            & Jup\,\&\,$\alpha$\,Tau  & $23^{\circ}42^{\prime}$ & $23^{\circ}43.0^{\prime}$ & $+1.0^{\prime}$ \\
Aug\,21     & Jup\,\&\,$\alpha$\,Tau  & $23^{\circ}40^{\prime}$ & $23^{\circ}42.0^{\prime}$ & $+2.0^{\prime}$ \\
            & $\alpha$ Ari\,\&\,Mira  & $26^{\circ}37^{\prime}$ & $26^{\circ}36.3^{\prime}$ & $-0.7^{\prime}$ \\
            & $\alpha$ Cet\,\&\,Mira  & $12^{\circ}51^{\prime \,(2)}$ & $12^{\circ}50.7^{\prime}$ & $-0.3^{\prime}$ \\
Sep\,1      & Jup\,\&\,$\alpha$\,Tau  & $23^{\circ}54^{\prime}$ & $23^{\circ}55.6^{\prime}$ & $+1.6^{\prime}$ \\ \hline
            &                         &                         &        mean & $1.6^{\prime}$ \\ 
            &                         &                         & std. dev.   & $1.5^{\prime}$ \\ \hline
\multicolumn{5}{l}{Meridian altitude measurements of Jupiter:} \\ 
\multicolumn{2}{l}{Date ({\it jul.})} & \multicolumn{2}{c}{separation}          & off- \\ 
1596        & Sect.       & Fabr.               & true     & set   \\ \hline
Aug 3       & 3.2 & $50^{\circ}2^{\prime}$ & $50^{\circ}5.1^{\prime}$   & $+3.1^{\prime}$ \\
Aug 3       & 3.1 & $50^{\circ}7^{\prime}$ & $50^{\circ}4.9^{\prime}$   & $-1.9^{\prime}$ \\
Aug 9       & 3.2 & $50^{\circ}7^{\prime}$ & $50^{\circ}9.2^{\prime}$   & $+2.2^{\prime}$\\
Aug 14      & 3.2 & $50^{\circ}12^{\prime}$ & $50^{\circ}10.2^{\prime}$ & $-1.8^{\prime}$ \\
Aug 17      & 3.2 & $50^{\circ}12^{\prime}$ & $50^{\circ}11.2^{\prime}$ & $-0.8^{\prime}$ \\
Sep 1       & 3.2 & $50^{\circ}4.5^{\prime}$ & $50^{\circ}4.4^{\prime}$ & $-0.1^{\prime}$ \\ \hline
            &                             &                         &      mean & $1.7^{\prime}$ \\ 
            &                             &                         & std. dev. & $1.1^{\prime}$ \\ \hline
\end{tabular}

\vspace{.2cm}
\noindent Remarks: (1) Fabricius was uncertain. \\
(2) Fabricius gave $12^{\circ}51^{\prime}$ or $12^{\circ}50.5^{\prime}$, see Sect. 3.2.
\end{table}

The offsets of all 17 separation measurements of Fabricius (Table 2) have a mean of $1.6^{\prime}$ with a
standard deviation of $1.5^{\prime}$.
The offsets of the six Jupiter altitude measurements have an offset of $1.7^{\prime}$ with a
standard deviation of $1.1^{\prime}$.
The offsets of all 23 measurements have an offset of $1.6^{\prime}$ with a
standard deviation of $1.3^{\prime}$.
Overall, we see a mean precision of $1.6-1.7^{\prime}$.

Given that the separations are specified to the very arc min (or even to half an arc min, Aug 21 and Sep 1)
and that separations between two stars did not change from day to day by more than one arc min, 
Fabricius could conclude on an internal precision of ca. 1$^{\prime}$ or better.

Then, on Aug 17, Fabricius noticed that the separations between Mira and $\alpha$ Ari had increased by 1$^{\prime}$
and that the separation between Mira and $\alpha$ Ceti had decreased by 1$^{\prime}$ -- both compared to three days earlier.
He did not conclude that the new object moved, but considered,
whether this was just a result of refraction (or maybe incomplete refraction correction).\footnote{Fabricius: `I think 
that the difference of the observations of the separations of the new star and the bright star of Aries
and the jaw (which was sometimes larger or smaller by one minute, as is clearly seen in the previous observations)
has its origin in the refractions, because I have sometimes observed these same separations faster,
and sometimes somewhat more roughly.
Because the jaw, lying closer to the horizon, will show a larger refraction and therefore a
smaller separation, even if perhaps the difference in separations between the comet and the bright star
cannot be excused/explained in this way, because the altitude of the bright star around that moment
was free of refraction -- I nevertheless think, that it 
[the difference] can be excused this way,
because the new star, which was made higher, had a smaller refraction and in the same extend then
a larger separation to the bright star' (Sect. 3.2).}
If Mira was at that time still more than $30^{\circ}$ above horizon
(which he mentioned for some other dates, but not here), he could consider 
refraction to be negligible following Brahe\footnote{Fabricius may have known Brahe's statement that refraction would be
negligible above $30^{\circ}$ -- 
Thoren (1990): `In Tycho's published table of refraction, the refractions vanished only above $45^{\circ}$
(Dreyer 1913, vol. II, p. 64); but Tycho frequently alluded to refractions being inconsiderable above $30^{\circ}$
(Dreyer 1913, vol. XI, pp. 15, 346, VI, p. 39)' (p. 230, note 27) and: `Tycho corrected for refraction simply
by choosing observations made above $30^{\circ}$, where the refraction, as far as he was concerned, was
negligible. But he always made formal corrections ...' (p. 231). Hence, when Fabricius observed Mira around culmination
at $31.5^{\circ}$, he could consider refraction as negligible following Brahe. In addition, the fact that Mira was indeed
at an altitude of $\ge 30^{\circ}$, confirms that our reconstructed observing times
are not far off.}, but -- on the other hand -- he considered refraction still
as probable cause for the apparent different separations.

Following the refraction approximation by Bennett (1982), refraction is $1^{\prime}$ at, e.g., an altitude of $45^{\circ}$,
and 1.6-1.7$^{\prime}$ at $32-30^{\circ}$ (Mira), respectively; or 0.6$^{\prime}$ 
at an altitude of $58^{\circ}$ like $\alpha$ Ari (when Mira is culminating).
The difference in refraction between Mira and $\alpha$ Ari can therefore amount to one arc minute,
but it would decrease the separation, because refraction causes astronomical objects to appear higher,
and the effect is larger at lower altitude than at higher.  

Since both $\alpha$ Ari and $\alpha$ Ceti are at higher altitude than Mira, their separations to Mira should
{\it both} decrease. This was not the case.
Also, on the next observing date, the separation between Mira and $\alpha$ Ceti was back at the 
previous value, while the separation between Mira and $\alpha$ Ari stayed one arc minute higher than original. 
Hence, refraction was probably not the cause for the variations on Aug 17 --
except, if Fabricius would have measured the separation between Mira with $\alpha$ Ceti on Aug 17 and 21 much
later and deeper on sky then earlier, and also the separation between Mira and $\alpha$ Ari on Aug 17
much earlier and higher on sky then otherwise, which he did not report at all.
For Aug 17 and 21, Fabricius mentioned to be somewhat uncertain about the separation between Mira and $\alpha$ Ceti.

Therefore, a slightly worse internal precision was probably the cause for these differences,
which may have been due to lower altitude or random measurement uncertainties.
Above, we noticed that the standard deviation of the separation measurement offsets is also around $1^{\prime}$.
Note that the separation measurement between Mira and $\beta$ Ceti 
(lowest altitude and largest separation to Mira), 
was not repeated by Fabricius.

Regarding the absolute position of Mira,
Fabricius gave the ecliptic coordinates of Mira in 1596 Aug at a 
longitude Aries $25^{\circ}47^{\prime}$ and a southern latitude of $15^{\circ}54^{\prime}$
(Sects. 3.1, 3.4, 3.5). This is also quite correct, the true values are:
longitude $25^{\circ}53^{\prime}$ and latitude $-15^{\circ}58^{\prime}$ (epoch of date),
i.e. just $4-6^{\prime}$ off.
 
In sum, the relative positional accuracy in Fabricius' separation measurements was already $\sim 1.3-1.7^{\prime}$,
and its absolute positional accuracy was $4-6^{\prime}$.
This also excludes other visible stars except Mira as identification of the `new star' of 1596.

Fabricius' data also show that the separation between Jupiter and Aldebaran was decreasing from Aug 3 to 21 --
only by the next (and last) measurement on Sep 1, the separation had increased (retrograde motion after stationarity).

Brahe (or one of his colleagues) also measured a few separations on 1596 Aug 8 in the morning
(Dreyer 1913, vol. XIII, pp. 36-37): \\
Jupiter to $\alpha$ Tau: $23^{\circ}56^{\prime}$, offset $2^{\prime}$ \\
Jupiter to $\alpha$ Peg: $55^{\circ}37~2/3^{\prime}$, offset $0.2^{\prime}$ \\
$\alpha$ Ari to $\alpha$ Tau: $35^{\circ}31^{\prime}$, offset $1^{\prime}$. \\
Hence, his measurements are at least of similar accuracy.

\subsection{On brightness and dating of the 1596 and 1609 maxima and Mira's period}

Of the 222 maxima of Mira recorded well on the AAVSO web page from 1660 to 2022, 
the mean brightness maximum lies at V=$3.05 \pm 0.54$ mag
(it is unlikely that the true maxima were strongly missed).
Mira varies in the V-band and visual band between 2nd mag (max. even 1.47 mag in 1779, see Fig. 2) and 10th mag 
(min. 10.4 mag) and in the B-band between 3.8 and 11.6 mag (see e.g. AAVSO long-term light curve). 
For earlier statistics, see Sigismondi et al. (2001).

In Sect. 3.6, we derived already the brightness of Mira during the 1596 maximum of $1.9 \pm 0.1$ mag --
based on the texts transmitted from Fabricius, see Table 1.
Hence, Fabricius discovered Mira at a particularly bright maximum.

Since 1660, only four of 222 recorded maxima were brighter than 2.0 mag
(regarding at least one, namely the brightest, data point):
1779 Nov 20 with 1.47 mag (Fig. 2), 
1906 Dec 13 with 1.60 mag, as well as 1997 Feb 19 and 2007 Feb 26 with 1.9 mag -- all from AAVSO (only V or visual band).
That two of these four -- or five, if we add the one in 1596 --
were more recent (1997 and 2007, plus the one in 2011 with a peak at 2.0 mag) may indicate
some secular variability, but could also be due to instrumental effects (CCD vs. photographic etc.).

That Fabricius caught a bright maximum has certainly facilitated its detection.
Fabricius could discover Mira, because he observed the close-by Jupiter shortly before and around
stationarity.\footnote{Clearly Jupiter at a separation of $\sim 20^{\circ}$ on Aug 13 ({\it greg.}),
not Mercury, as given in some previous publications on Fabricius' discovery of Mira --
Kepler mistakenly wrote in his {\it Optics} that Fabricius would have measured the
separation between the new star and Mercury, see e.g. Hatch (2011, p. 167).}
As already mentioned (Sect. 3.6), the broad Mira maximum ranges from its first detection 
on Aug 13 to Aug 31 ({\it greg.}), after which Fabricius noticed that it got fainter; the middle is Aug 22 ({\it greg.}).
For 1609, neither brightness is given nor a maximum date can be derived from the text (Sect. 3.6).
For the remainder of this paragraph (Sect. 4.2), we will use only Gregorian dates.

\smallskip

{\bf
Can we constrain the brightness of Mira in 1609? 
}
Since Fabricius mentioned that the nearby horizon was not the problem for losing
Mira after Mar 4 (Sect. 3.5, letter iii), we can assume that he observed it at low altitude, at the end below
$10^{\circ}$, say in the middle of the astronomical twilight (Sun $15^{\circ}$ below horizon).
In the evening of the re-detection by Fabricius on 1609 Feb 15, the sky brightness at the position of Mira
at that time, as seen
from the location of Fabricius as very experienced observer\footnote{We use 
{\url http://www.cleardarksky.com/others/BenSugerman/ star.htm}
with the following inputs: Longitude and latitude for Osteel, elevation 0 meter (sea level), 
1609 Feb 15, temperature $50^{\circ}$ Fahrenheit, relative humidity $40\%$, 
for an experienced observer with Snellen ratio 1, experience 10 (high),
and age of 45 years, at the sky location of Mira, extinction coefficient 0.3 mag/airmass.}
is $\sim 18.4$ mag per arcsec$^{2}$, and the limiting visual magnitude is then $4.2 \pm 0.4$ mag.
Hence, Mira would be detectable if at least as bright as 3.8-4.6 mag (or at least $\sim$3.0 mag for $3\sigma$ above the limit).
Note that Mira suffered from 0.3 mag atmospheric extinction at this moment.
Hence, when Fabricius detected Mira first on 1609 Feb 15, it was at least in the range
of 3.5-4.3 mag or brighter (or even $\sim$2.7 mag for $3\sigma$ above the limit).

On the date of the last detection, Mar 4, the limiting magnitude was $4.4 \pm 0.7$ mag,
so that Mira would need to have been, after correction for 0.6 mag extinction, at 3.1-4.5 mag 
(or 1.7 mag or brighter at $3\sigma$, also corrected for extinction).
In the next ten days, the limiting visual magnitude degraded
to $0.0 \pm 1.6$ mag on Mar 14 (first cold air date, i.e. probably clear sky, extinction 1.0 mag) --
always for the Sun being $15^{\circ}$ below horizon
and Mira being at least a few degrees above horizon;
Mira was never observed nor expected to be that bright.

\smallskip

{\bf On the period of Mira:}
We will consider different period determinations, namely two closest to Fabricius' time,
one from most recent data, and one over the longest possible time interval.
From the maximum by Fabricius on 1596 Aug 22 (JD 2304221)
to Hevelius' maximum on 1660 Nov 1 (Fig. 4, JD 2327668 $\pm$ few days on AAVSO and Guthnick 1901), one can obtain a period of 330.24 days.
From the two maxima observed by Hevelius in 1660 and 1678 (1678 maximum on JD 2334253 $\pm$ few days), 
one can obtain a period of 329.25 days.
The period of Mira was most recently given as 330.2 days on VSX for the epoch 2004 to 2023.
From the 18 period determinations in the literature from 1660 to 1981 (Hoffleit 1997) plus the most recent one (VSX),
we obtain a mean of 331.0 days with a standard deviation of 3.5 days.

If we extrapolate from the Mira maximum on, say, 1596 Aug 22 with those four periods to
1608 and 1609, we obtain:
\begin{itemize}
\item With period from Fabricius 1596 to Hevelius 1660 (330.24 days), we expect maxima ca. 1608 May 24 and 1609 Apr 19.
\item From the two Hevelius maxima (329.25 days), we expect maxima ca. 1608 May 11 and 1609 Apr 5.
\item With the VSX period (330.2 days), we expect maxima ca. 1608 May 24 and 1609 Apr 19.
\item With our long-term mean (331.0 days), we expect maxima ca. 1608 Jun 3 and 1609 Apr 30.
\end{itemize}
The first and third period are almost identical.

From the light curves shown in Figs. 2-4 (in particular the well-covered light curve for 2011,
for which we have fitted a mean light curve, Fig. 3 lower left), we can derive a decrease of 1 mag in $23 \pm 2$ days after the peak
(3-day bins with box-car smoothing).
In 2011, the brightest data point was 2.0 mag, while the mean brightness within 3 days of the peak is $2.4 \pm 0.4$ mag.
(The total range in these days is even 4.3 mag with three faint outliers. The relatively large scatter in brightness per day,
even though obtained by CCD photometry, might be due to different observational setups, data reduction techniques, and standard stars.)

Since early July 1608, Mira should have been visible before sunrise,
if it would have still been bright enough after the maximum around 1608 May 11 to Jun 3.
Then, with its typical decrease, Mira might have been $\sim 1.9 \pm 0.5$ mag fainter by early July 1608 
than at the 1608 maximum. However, it was not detected in 1608 by Fabricius, 
even though he probably expected to detect it 
(see Sect. 3.6 in Mira theory).
In Sect. 4.4 below, we will show that a star as red as Mira can hardly be detected by the naked eye
if fainter than ca. 4.5 mag.
Therefore, the maximum in 1608 was probably fainter than ca. $2.6 \pm 0.5$ mag.

The observations by Fabricius 1609 Feb 15 to Mar 4 were still a few weeks before the maximum expected from its period,
namely 1609 Apr 5-30.
For a brightening by 1.5 mag in $23 \pm 2$ days from the value on Mar 4 (at least 3.1-4.5 mag), we would expect 1.0-2.4 mag already on Apr 5.
This would indicate another relatively bright maximum in 1609 --
again facilitating the detection by Fabricius.
For the brightest recorded maximum, 1779, Mira was around 2nd mag or brighter (peak at 1.47 mag) for some 55 days (Fig. 2).
(Even if the 1609 maximum was already on Mar 14, this would give a time interval of $\sim$4587 days
for 14 periods, i.e. a period of $\sim$327.6 days; this is within up to 3.3 days of the above considered periods.)

The periods determined from both the most distant time (Fabricius' maximum in 1596 to Hevelius in 1660) 
and the most recent data (VSX for 2004-2023) are both $\sim$330.2 days.
If we extrapolate from the well-covered maximum in 2007 (ca. JD 2354154.5 $\pm$ few days) to 1596 with this period,
the maximum is expected only ca. 23 days later;
such an offset is well within the range observed in the last four centuries (O$-$C being in the range of $-40$ to $+51$ days,
plus one outlier at $+89$ days, see Hoffleit 1997).
We see no evidence for a long-term secular trend (or jump) in the period, 
i.e. a stable pulsation.
(It would be beyond the scope of this paper to study the period variability from year to year over
four centuries by elaborated fits to all data.)

\subsection{Color index of Mira 1596 Aug-Oct}

The color statements by Fabricius on Mira 
can be quantified as B$-$V$\simeq$1.3-1.4 mag, see Sect. 3.6, Table 1.

Can we estimate the B$-$V color index of Mira in bright maxima and during the subsequent faintening also alternatively,
e.g. with modern CCD data?

There are 359 cases in the AAVSO data table for Mira (in the CCD era from 2003 to 2023 Apr 6),
where the same observer has taken both B- and V-band data in the same night, 
mostly obtained by the observers G. Samolyk, J. Centala, and J. DeYoung.
When Mira is at V$\simeq$2.0 to almost 3.5 mag, we find B$-$V$\simeq 1.3 \pm 0.1$ mag --
consistent with the quantification of the color and brightness statement given by Fabricius.

For Mira-like pulsating variables, it is known that the apparent brightness (V-band or filterless) is strongly
correlated with the effective temperature: When the star is brightest, it is also hottest.
Hence, we can expect to see a correlation between the V-band/visual magnitude and the B$-$V color index in the AAVSO data.

We show in Fig. 2 two periods since the CCD era well covered in both
the visual light curve (left) and the B$-$V color index (right) -- the maximum in 2011 (bottom) and the minimum and maximum in 2007 (top);
in 2011, Mira reached a relatively bright peak of 2.0 mag (visual), 
the mean maximum was $2.4 \pm 0.4$ mag.
The B$-$V color indices indeed show minima near the brightness peaks, namely B$-$V$\simeq 1.25$ mag at 
a mean maximum of 2.37 mag (observer Samolyk) and $\simeq 1.38$ mag (Centala) in 2011,
as well as B$-$V$\simeq$1.5-1.6 mag at $\sim$3.5 mag in 2007.
Below V$\simeq$3.5 mag, the color index B$-$V increases faster (see Fig. 2, upper right part).
A value of B$-$V$\simeq 1.3 \pm 0.1$ mag seems appropriate for a bright maximum (like in 2011 and 1596). 

In Fig. 2, we see also that brightness and color evolution are faster before the maximum.

\begin{figure*}
\centering
\includegraphics[angle=270,width=17.5cm]{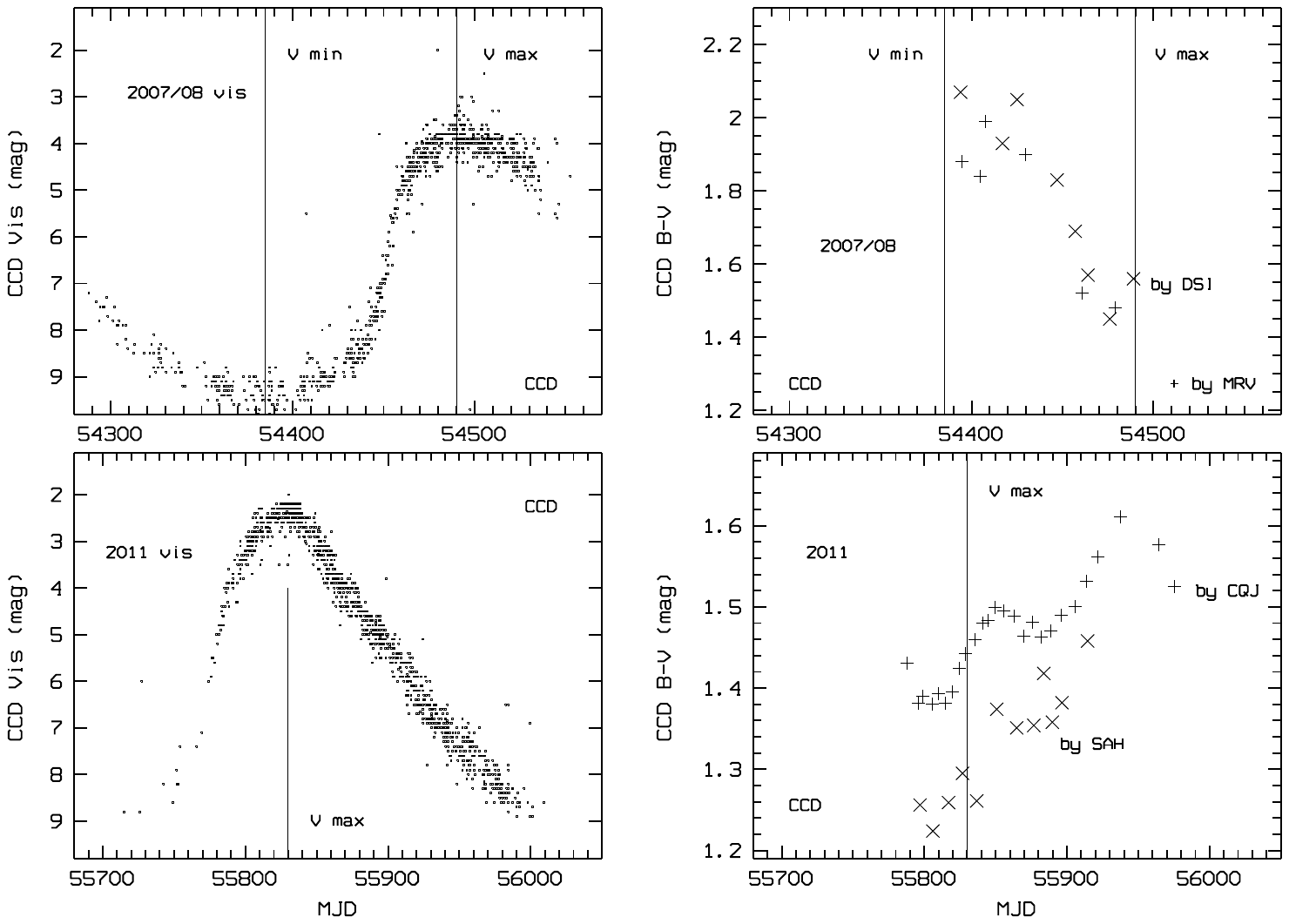}
\caption{{\bf Mira -- Visual magnitude and B$-$V color index for periods 2007 and 2011.}
For the two periods in 2007 (top) and 2011 (bottom),
we show the visual light curves (left) and the B$-$V color indices for the same time (right) --
all are CCD data from AAVSO. 
We indicate maxima and minima by vertical lines.
The B$-$V CCD data from particular observers (only those with both B and V from the very same night)
show small offsets, probably due to different instrumentation, comparison stars, etc.
AAVSO Observers: G. Samolyk (SAH), J. Centala (CQJ), G. Di Scala (DSI), and R. Modic (MRV).
We see that Mira is bluer when brighter, as expected and known.}
\end{figure*}

\subsection{Down to which magnitude could Fabricius follow Mira?}

We can now consider until which brightness and color index Fabricius may have observed Mira --
all dates on the Gregorian calendar.
In 1609, his last detection of Mira was only 17 days after the first (1609 on Feb 15 to Mar 4), 
then Mira became invisible due to conjunction with the Sun.
In 1596, Fabricius discovered Mira on Aug 13 
and reported a faintening after Aug 31, so that we see a broad maximum (as in modern CCD
light curves, Figs. 2 \& 3). Then, he
followed it until disappearance shortly before or during Oct 12-26 (Sect. 3.6).

\begin{figure*}
\centering
\includegraphics[angle=270,width=17.5cm]{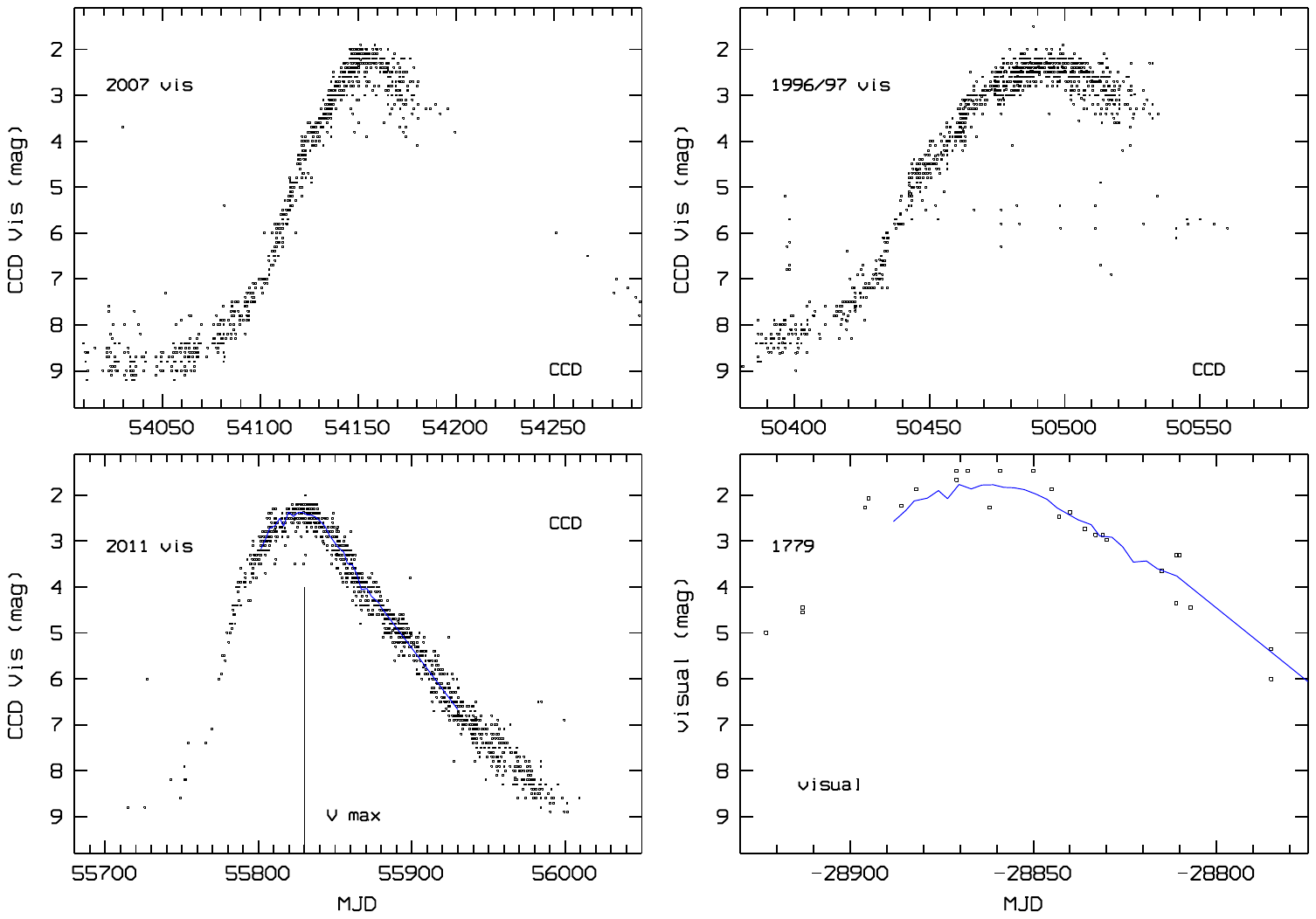}
\caption{{\bf Mira -- light curves for four bright maxima.}
We show the visual light curves for Mira for three bright maxima and the
well-covered one in 2020 -- all from AAVSO:
2020 (top left, CCD), 2007 (bottom left, CCD), 1996/97 (top right, CCD), and 1779 (bottom right),
the latter obtained by Johann Bode, William Herschel, Johann Schroter, and Pehr Wargentin 
(Guthnick 1901, pp. 130-131, see also AAVSO).
In the lower panels, we show our best fit to the main part of the light curve of 2011
(around maximum and early decrease) as blue curve (3-day binning with box-car smoothing) --
in the lower right panel shifted to overlap with the mean maximum of 1779.
}
\end{figure*}

Fabricius reported a brightness corresponding to $1.9 \pm 0.1$ mag (middle of the broad maximum ca. Aug 22), 
Table 1 and Sects. 3.6 and 4.2;
since Mira was well visible at night-time in 1596, 
he should have been able to follow it for more than 3 months (faintening by 1 mag in $23 \pm 2$ days), 
if visible down to ca. 6th mag.
Extinction should not have been the problem for detection, as it is only a few tenth of mag at Mira's meridian passage as seen from Frisia.
Mira would have reached 6th mag after $94 \pm 8$ days, i.e. in the last third of November.

\begin{figure*}
\centering
\includegraphics[angle=270,width=17.5cm]{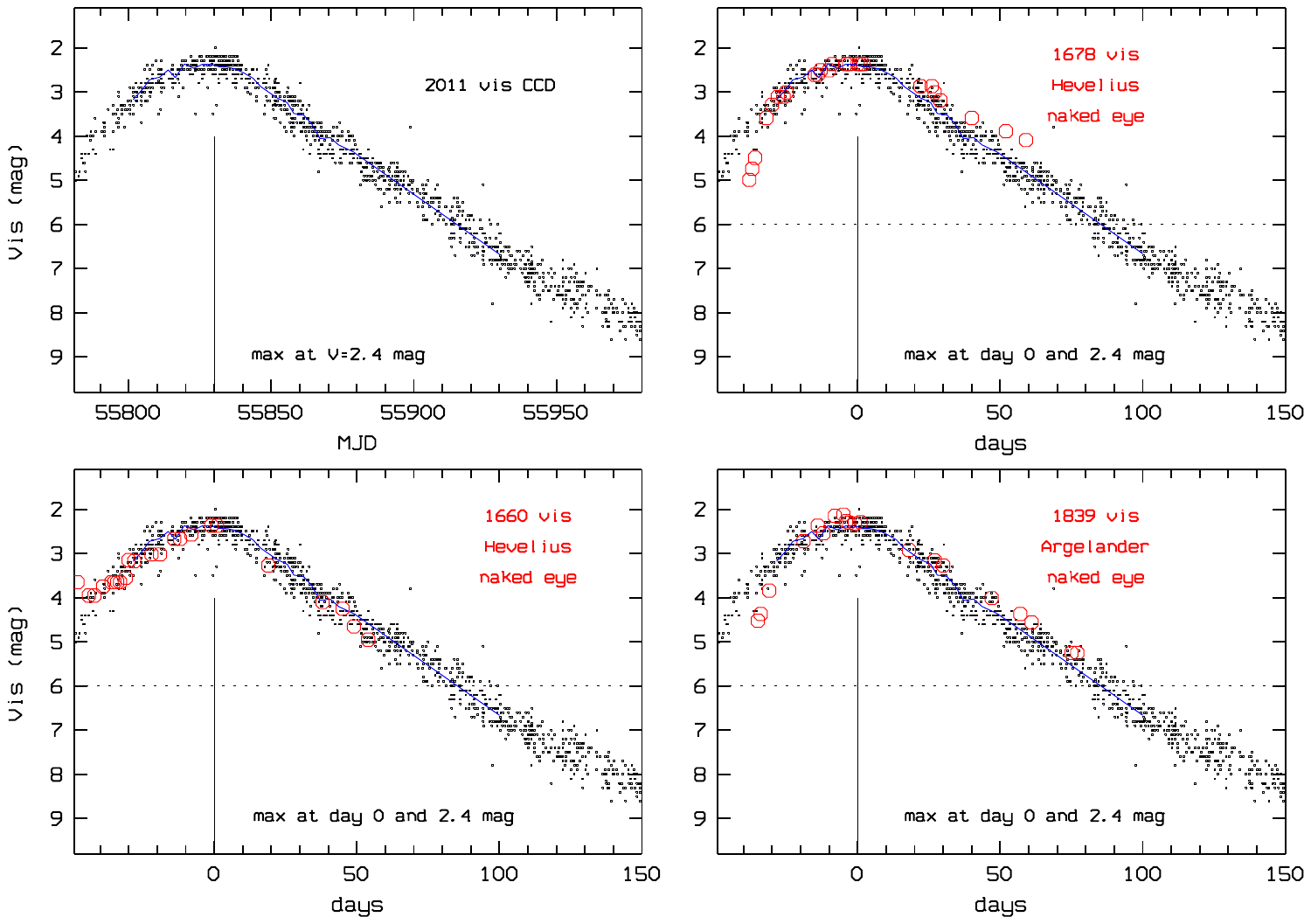}
\caption{{\bf Mira -- naked-eye observations of the faintening after four bright maxima.}
We show the visual light curves for Mira after the bright maxima in 2011 (modern
CCD data from AAVSO, upper left, black dots, 
peak at 2.0 mag, mean maximum at $2.4 \pm 0.4$ mag)
and then for 1660 (lower left, 
peak 2.07 mag) 
and 1678/79 (upper right, 
peak 2.23 mag)
from Hevelius and for 1839 (lower right, 
peak 2.12 mag) 
from Argelander -- data from Hevelius and Argelander as red circles.
Data are taken from Argelander (1869), Guthnick (1901), and AAVSO -- omitting data with high uncertainty
due to the close bright moon, unstable atmosphere, etc. (see Guthnick 1901).
Except in the upper left, we set the maxima 
to 2.4 mag 
at day 0 to plot the historical light curves over the modern one from 2011.
}
\end{figure*}

The reported disappearance is dated once to soon after St. Michael (Sep 29 {\it jul.} = Oct 9 {\it greg.}),
most likely shortly before or during the battles in Hungary, Oct 12-26, see Sect. 3.6.
Full moons were on 
1596 Oct 5/6 and 
Nov 5/6. 
Hence, we seem to have a discrepancy of up to 
five weeks (ca. Oct 20 to last third of Nov).

The wavelength-dependence of the sensitivity of the human eye is different during the day
(cones with maximal sensitivity around 550 nm) and night (rods with maximum around 500 nm) --
it is best sensitive in the yellow-green wavelength range.
The naked-eye limit has been estimated by Schaefer (1990) for typical stars
for a typical sky surface brightness of 20.9 mag/arcsec$^{2}$ at the zenith to be V=$6.0 \pm 0.5$ mag
(similar in Crumey 2014).

{\bf Naked-eye limit for very red stars:}
Let us now estimate the dependence of the naked-eye limit for stars on their B$-$V color index.

The spectral sensitivity of the human eye varies with illumination. 
In bright light, the eye is more sensitive to the red part of the electromagnetic spectrum, 
whereas the fully dark-adapted eye is more blue-sensitive. 
The blue shift of the so-called scotopic vision of the human eye compared to the photopic vision 
during the day can be explained by the fact that at dusk or dawn only highly scattered sunlight illuminates the sky, 
which is also affected by the absorption 
of ozone in the green part of the electromagnetic spectrum. The spectral response of the human eye as a function of illumination from 
Williamson \& Cummins (1983) and the transmission of the V-band filter as defined by Bessell (1990) are shown in the middle panel of Fig. 5.

Since the spectral flux density of stars varies with wavelength and the spectral response of the human eye 
differs from the spectral transmission of the V-band filter, 
the visual limiting magnitude must also depend on the color of the observed star. 
Since the fully dark-adapted human eye has a higher response in the blue compared to 
the transmission of the V-band filter, the visual limiting magnitude should decrease 
with increasing color index B$-$V of the star. The spectral flux density of a 
star $F_\lambda$, the scotopic spectral response of the human eye $r$ and the spectral 
transmission of the V-band filter $t$ can be used to calculate the fluxes, 
that can be detected by the human eye:
\begin{equation}
F_{\rm sco} = \int r(\lambda) \cdot F_\lambda(\lambda) \cdot d\lambda
\end{equation}\newline
or through the V-band filter:
\begin{equation}
F_{\rm V} = \int t(\lambda) \cdot F_\lambda(\lambda) \cdot d\lambda
\end{equation}\newline
The flux densities $F_\lambda$ are taken from the Pickles (1998) library of stellar spectra, the intrinsic B$-$V colors 
and corresponding spectral types of main sequence stars from Pecaut \& Mamajek (2013).

\begin{figure}
\resizebox{\hsize}{!}{\includegraphics{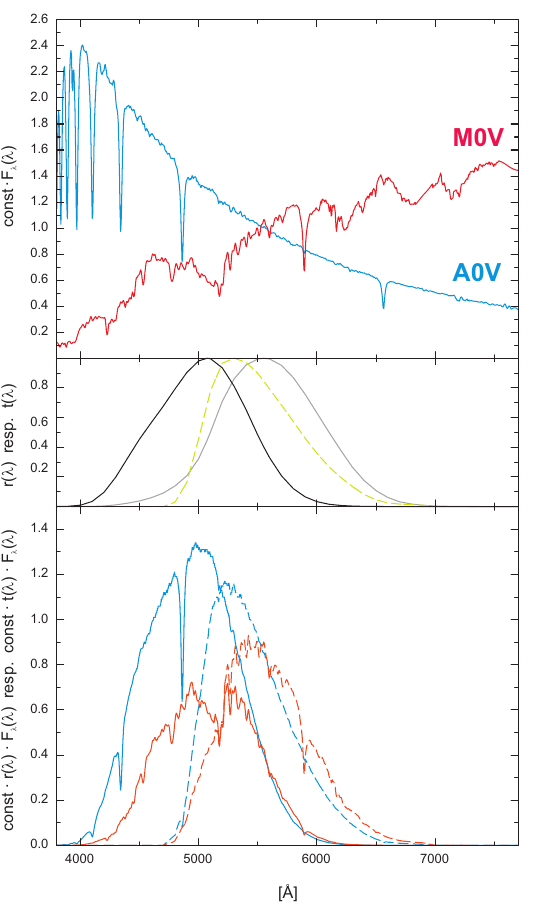}}
\caption{\textbf{Top:} The spectral flux density $F_{\lambda}$ of an A0 (blue spectrum) and M0 (red spectrum) main sequence star. 
\textbf{Middle:} The spectral response $r$ of the scotopic (black curve) and photopic (grey curve) vision of the human eye, 
and the transmission of the V-band filter (dashed green curve). 
\textbf{Bottom:} The spectral flux density of an A0 (blue spectra) and a M0 main sequence star (red spectra) 
as seen by the fully dark-adapted human eye (solid spectra) or 
measured through the V-band filter (dashed spectra).}
\end{figure}

The derived flux ratio $F_{\rm sco} / F_{\rm V}$ for main sequence stars of different spectral types are summarized in Table 3. 
For white stars (like Vega with B$-$V=0.0 mag),
the flux detected in scotopic vision is about 40\,\% higher than through the V-band filter. 
In contrast, for late-type stars, e.g. a M0V star with B$-$V=1.4 mag, it is about 13\,\% lower. 
This leads to a change in the visual limiting magnitude of $\Delta V_{\rm Limit} = 2.5 \cdot \log(0.873/1.406) = -0.52$ mag between observing an 
A0 and a M0 main sequence star. The visual limiting magnitude for very red stars can be up to 0.6 mag lower than that for white stars. 
This means that late-type stars must be about a factor of two brighter in the V-band than early-type stars 
in order to still be visible to the naked eye in the dark night sky.

\begin{table}[h]
\caption{Spectral type (SpT), intrinsic B$-$V color, derived flux-ratio $F_{\rm sco} / F_{\rm V}$, 
and estimated change in visual limiting magnitude 
$\Delta V_{\rm Limit}$ for main sequence stars, as a function of their color.}
\begin{center}
\begin{tabular}{lccc}
\hline
SpT & B$-$V [mag] & $F_{\rm sco} / F_{\rm V}$ & $\Delta V_{\rm Limit}$ [mag] \\
\hline
A0V & 0.000         & 1.406                     & \,\,\,\,0.00 \\
F0V & 0.295         & 1.262                     & $-$0.12 \\
F5V & 0.440         & 1.198                     & $-$0.17 \\
G0V & 0.595         & 1.161                     & $-$0.21 \\
K0V & 0.816         & 1.084                     & $-$0.28 \\
K3V & 0.990         & 1.014                     & $-$0.35 \\
K5V & 1.150         & 0.911                     & $-$0.47 \\
M0V & 1.420         & 0.873                     & $-$0.52 \\
M4V & 1.650         & 0.837                     & $-$0.56 \\
M5V & 1.830         & 0.813                     & $-$0.59 \\
M6V & 2.010         & 0.793                     & $-$0.62 \\
\hline
\end{tabular}
\end{center}
\end{table}

The naked-eye detection limit of $6.0 \pm 0.5$ mag was calculated for typical star colors and zenith observations
(e.g. Schaefer 1990).
For observations carried out at a given (non-zero) zenith distance $z$, due to the atmospheric extinction $k$,
the detection limit is reduced by $\Delta V_{\rm limit} = -k \cdot (1/\cos(z) - 1)$.
At the location of Fabricius, Mira has a culmination altitude of $31.5^{\circ}$,
so that it can be observed for some time at a zenith angle of $z \simeq 60^{\circ}$ (airmass ca. 2) --
e.g., in August in the early morning (when Fabricius observed), in mid October also around midnight (when the sky is even darker).
For an atmospheric extinction of k$\simeq$0.3 mag/airmass,
which should be applicable to Fabricius' site near the coast of the North Sea,
the above equation yields a reduction of the detection limit by $\sim$0.3 mag at a zenith distance of about 60$^{\circ}$.
Furthermore, to the increased sky surface brightness at this zenith distance, the brightness limit is reduced by 
further $\sim$0.3 mag compared to the zenith (Garstang 1989).
Together, this reduces the limit by $\sim$0.6 mag.
For the sky location of Mira at airmass 2 (Fabricius), we then arrive at a limit of $\sim 5.4 \pm 0.5$ mag --
for stars with typical color.

Now, the median B$-$V color index for all stars in the Bright Star Catalogue (Hoffleit \& Jaschek 1991) down to the above limit
of V=5.4 mag is B$-$V=0.44 mag, which corresponds to an F5V star.

Since the middle of the broad maximum of Mira at $1.9 \pm 0.1$ mag (1596 Aug 22),
it faintens by ca. 1 mag in $23 \pm 2$ days, i.e. to $4.47 \pm 0.31$ mag Oct 20,
when Fabricius lost it. At such a brightness,  
it had a color index of B$-$V$\simeq 1.6-1.7$ mag 
(by analogy to 2011, see Fig. 2, upper panels: e.g., B$-$V$\simeq$1.6 mag when at V$\simeq$4.0 mag).
Then, the difference in the detection limit (offset) between the typical F5V star with B$-$V=0.44 mag and Mira when disappearing
to the naked eye (B$-$V=1.6-1.7 mag), according to Table 3, is $\sim$0.4 mag.
Hence, the detection limit for the location of Mira, when at B$-$V=1.6-1.7 mag, is reduced to $5.0 \pm 0.5$ mag.
If we accept Fabricius as a particularly good observer, we would expect his limit towards the fainter end of this range,
i.e. almost 5.5 mag,
but see also below for further evidence as to when very red stars become invisible.
Note that he did not report to have detected Uranus, which was located in the same general field amongst Jupiter,
Aldebaran, Hamal, and Mira, close to $\mu$ Psc, with V=5.7 mag (reduced to 6.15 mag due to airmass 1.5 for k=0.3
mag per airmass during his observations of this field) in Aug and Sep 1596.

The main profession of David Fabricius was being a protestant pastor, so that he might have observed
the sky mostly in the early morning and the early evening, but not so much in the middle of the night;
indeed, all his observations of Mira in AD 1596 (Sect. 3), 
where he specified roughly the observing time,
were obtained in the morning or evening (nautical and civil twilight). 
If he would have observed also around Oct 9 (St. Michael on greg. calendar) only in the early morning, i.e. around
the time when the Sun was only $11.5^{\circ}$ below horizon, Mira would have had an airmass of $\sim$5,
i.e. invisible for naked-eye observations.
Since he still spotted Mira, he must have observed earlier in the night
(now without instruments as he stated in his letter to Brahe, Sect. 3.2).

Fabricius would need to have searched for Mira (in October) at an altitude of $\ge 20^{\circ}$, i.e. an airmass of $\le 3$,
then (for k=0.3 mag/airmass) his detection limit would be reduced by additional $\sim$0.5 mag due to 
increased extinction and sky background, i.e. to $4.5 \pm 0.5$ mag -- consistent with the brightness of Mira we just
estimated for the second half of October 1596.

\smallskip

{\bf Further evidence:}
Finally, we consider further empirical evidence for the effect that faint red star are more difficult to detect:

(a) Deep red stars and their naked-eye visibility: Among the six brightest stars with V=3-6 mag and B$-$V$\ge 1.7$ mag
(in the Tycho-Hipparcos star catalog, ESA 1997), 
only the 2nd and 3rd brightest are enlisted by Ptolemy
(we consider only the Almagest here, where the star identifications are based on coordinates, 
while other reconstructions of constellations and asterisms, like the Chinese, are too uncertain for very faint stars):  \\
the brightest is $\sigma$ CMa (V=3.47 mag, B$-$V=1.73 mag) -- not in the Almagest, \\
the 2nd brightest is {\it o}$^{1}$ CMa (V=3.84 mag, B$-$V=1.76 mag), Almagest CMa 13; \\ 
the 3rd brightest is $\mu$ Cep (V=4.02 mag, variable 3.6-5.0 mag on AAVSO, B$-$V=2.23 mag), Almagest Cep 12 
-- close to zenith; \\
the 4th brightest is 119 Tau (V=4.36 mag, B$-$V=2.03 mag), \\
the 5th is BE Cam (V=4.64 mag, B$-$V=1.86 mag), and \\
the 6th is {\it o}$^{1}$ Ori (V=4.74 mag and B$-$V=1.75 mag), \\
none of them in the Almagest. \\
Even though the Almagest is not complete at a brightness fainter than 4 mag,
this may indicate that the naked-eye limit for deep red stars (B$-$V$\ge 1.7$ mag) is around 4-4.5 mag --
and slightly brighter for stars with B$-$V=1.5-1.7 mag. 

(b) Mira today: The experienced naked eye variable star observer Frank Vohla could follow Mira after one of the recent maxima 
by the naked eye down to an estimated visual 4.5 mag on 2019 Dec 4 in Altenburg, Germany, and Dec 5 in Leipzig Altlindenau, Germany
(F. Vohla, priv. comm.), when it is listed at a visual magnitude of $\sim 4.0$ mag on AAVSO. 
According to Fig. 2, it would have B$-$V$\simeq 1.6-1.7$ mag at a brightness of 4.0-4.5 mag.

(c) The `garnet star' $\mu$ Cep: The above mentioned star $\mu$ Cep is deep red (V=4.02 mag, B$-$V=2.233 mag, 
M2 Ia, variable from V=3.1-5.4 mag),\footnote{The star was named `garnet star' by W. Herschel, 
who observed it with a telescope, but this neither confirms that he saw the color 
of the star by the unaided eye, nor is it possible to quantify a possibly meant color index.}
but was never reported to be seen in reddish color by the unaided eye -- except maybe due to extreme chromatic scintillation. 

(d) Mira observations by Hevelius (1611-1687, Gdansk, Poland) and 
Argelander (1799-1875, Bonn, Germany): As reported by Argelander (1869), Hevelius (1660 and 1678/79) 
and he himself could detect Mira by the naked eye up to almost three months after
particularly bright maxima:\footnote{While some magnitudes given next from Hevelius and Argelander,
like also those from 1779 (max V=1.47 mag), seem to indicate a high precision, as given to the
2nd digit, i.e. to one hundredth of a mag, they are in fact not that precise: These brightness values were obtained
on a different scale and converted to our current Pogson magnitude scale (Pogson 1856) later 
and thereby got an apparent precision to the 2nd digit (see Guthnick 1901 for Mira).} \\
Hevelius observed from 1660 Nov 2 ({\it greg.}) at 2.07 mag to 1661 Feb 2.5 at presumably 6.6 mag.
However, Bullialdus in Paris obtained on 1661 Feb 23.5 an uncertain 6.1 mag (Guthnick 1901, p. 125).
Then, the nine observations from 1660 Dec 31.5 at 5.35 mag to 1661 Feb 2.5 are uncertain
according to Guthnick (1901, p. 127, who also consulted the work by Hevelius) 
due to the close bright moon and/or other unspecified reasons,
so that the last unproblematic measurement was on 1660 Dec 25.5 given at 4.65 mag. \\
Hevelius observed also from 1678 Nov 19 at 2.23 mag to 1679 Feb 6 at 6.0 mag.
However, the four observations from 1679 Jan 19 at 4.35 mag to Feb 6 at 6.0 mag are uncertain
according to Guthnick (1901, p. 127) due to the close bright moon and/or other unspecified reasons,
so that the last unproblematic measurement was on 1679 Jan 15 given at 3.95 mag. \\
Argelander observed from 1839 Oct 8.5 at 2.12 mag to 1840 Jan 5.5 at 6.0 mag.
However, on 1839 Jan 2.5 and 5.5, the observations were taken with an opera glass (magnification by a factor two) 
or 24-inch comet searcher at unstable atmospheric conditions
according to Guthnick (1901, p. 135), so that the last unproblematic naked-eye measurement was on 1839 Dec 29.5
given at 5.25 mag. \\
For Hevelius at Gdansk, Poland, Mira culminated 
at $31^{\circ}$,
and for Argelander at Bonn, Germany, Mira culminated 
at $35.5^{\circ}$,
so that a similar limit applies for the detection by the unaided eye as calculated before
for Fabricius, namely $5.0 \pm 0.5$ mag for stars as red as B$-$V$\simeq 1.6-1.7$ mag at airmass $\sim$2 mag --
and indeed, two of these three measurements lie within the $1\sigma$ range (all 3 within the $2\sigma$ range):
The apparent limit of Hevelius (aged 68 years in 1678/79), 
is slightly brighter than the general limit, 
and the one by Argelander (aged 40 years in 1839/40) lies at the faint end of the range --
possibly just due to their age difference.
Argelander (1869, p. 14) also wrote that Mira was rarely visible to the unaided eye for more than two months after the maximum.

After the above three maxima (AD 1660, 1678/79, 1839), 
Mira was detected with the unaided eye until 54 to 83 days,
so that the brightness should have decreased by only $\sim 2.3-3.6$ mag (1 mag in $23 \pm 2$ days);
then, according to Fig. 2, we expect the color index to be B$-$V$\simeq 1.4-1.7$ mag.

In sum, we can conclude that around ca. 4.5 mag, a deep red star like Mira becomes invisible for the naked eye.
This would be consistent with Fabricius having seen Mira at $\sim 1.9 \pm 0.1$ mag since Aug 13,
faintening noticeable after Aug 31 (maximum ca. Aug 22),
and then visible for about two months until disappearance ca. Oct 12-26.
If the maximum was around 1596 Aug 22 and if Mira can then be observed for 54-83 days
(Hevelius and Argelander), then Mira would have been visible until 1596 Oct 14 to Nov 12,
which is in good accord with the transmissions from Fabricius (Sect. 3.6).
And 54-83 days  
after a maximum at $1.9 \pm 0.1$ mag (Fabricius 1596), the expected brightness would lie at ca. 4.2-5.5 mag.

\smallskip

In the next few years, we plan to monitor Mira both with simultaneous V- and B-band CCD photometry
and naked-eye observations to improve our knowledge on naked-eye detection of stars when they get redder
(and the intercomparison of CCD and naked-eye data).

\section{Summary}

Fabricius observed Mira on 1596 Aug 13, 19, 21, 24, 27, 31 as well as in Sep and Oct
and in 1609 on Feb 15, 22, March 1, 2, and 4.
All dates in this section are Gregorian.

We summarize here our main results and conclusions:
\begin{enumerate}
\item Mira was not part of Hipparchos' asterism Cetus. 
His Greek term `lophias' means `mane' and points to $\xi^{1}$ Ceti (Sect. 2\,i).
\item The guest star seen AD 1592-1594 in Korea in Tiancang was not Mira, 
in particular because its given position is strongly inconsistent with Mira 
and not located within $\sim 0.5^{\circ}$ of some other visible star (Sect. 2\,ii).
\item The records of the `guest star' seen AD 1070 Dec 25 in China in Tianjun (and lunar mansion right ascension range Lou) 
pertains much more likely to a comet than Mira -- given additional reports from Armenia and Germany for winter AD 1070/71 (Sect. 2\,iii).
\item Mira probably was part of the Classical Chinese asterism Tianjun based on old maps and modern identifications, see Fig. 1,
and would then not be reported as guest star (Sect. 2\,iv).
\item All relevant texts on Fabricius' discovery of Mira are presented with our literal translations to English
(Sects. 3.1-3.5) and discussed in Sect. 3.6; all essential data derived from the texts are compiled in Table~1.
\item Fabricius measured the position of Mira (relative to other stars) with an accuracy of $\sim$1.6-1.7$^{\prime}$ (Sect. 4.1).
\item We revise the brightness of Mira at the 1596 maximum (previously 2.80 mag, e.g. on AAVSO):
Fabricius specified the brightness as `slightly brighter than the bright [star] of Aries',
i.e. $\alpha$ Ari (V=$2.020 \pm 0.004$ mag), so that Mira had V$\simeq 1.9 \pm 0.1$ mag as maximum in 1596.
This is consistent with the also given `2nd mag' on the Almagest nomenclature (Sects. 3.6 \& 4.2). 
(For 1609, the maximum was probably also brighter than given on AAVSO, see Sect. 4.2.)
\item Fabricius described the color of Mira for the 1596 maximum as `red' and `like Mars' as well as `pale red like Saturn and Mars';
and he did not compare Mira in color with the nearby and otherwise mentioned Hamal and Aldebaran;
this means a color index of B$-$V$\simeq$1.3-1.4 mag (Sect. 3.6).
CCD observations of Mira on the AAVSO database indicate that Mira has such a color index when at V=2.0-3.5 mag (Sect. 4.3). 
\item After its discovery on Aug 13 and a broad maximum (middle: Aug 22), 
Mira began to fainten after Aug 31, and Fabricius lost it in mid/late Oct 1596 (Sect. 3.6)
-- probably at ca. 4.5 mag
(Sect. 4.4). This is consistent with modern light curves, where
Mira faintens by ca. 1 mag in $23 \pm 2$ days after the maximum (Figs. 2-4).
\item Fabricius re-detected Mira 1609 Feb 15 to Mar 4. 
We estimate the sky brightness and magnitude limit for these dates and 
find that these observations are again close to a relatively bright maximum (Sect. 4.2).
He probably also searched for Mira already in early July 1608, given the prograde motion of the nearby Jupiter 
(see Sects. 3.6 and 4.2), roughly 1-2 months after the maximum -- 
and since he did not detect Mira, this maximum must have been fainter than ca. $2.6 \pm 0.5$ mag.
\item Mira's period is 330.2 days in the oldest data (from the maxima observed by Fabricius 1596 and Hevelius 1660)
and in the newest data (VSX for 2004-2023). Also, from the well-covered maximum in 2007 extrapolated to 1596 with a
period of 330.2 days, we arrive at the 1596 maximum within less than one month. Hence, we see no evidence for a 
long-term trend in the Mira period or phase (Sect. 4.2).
\item We show that the detection limit of the naked eye for very red stars (B$-$V$\simeq$1.6-1.7 mag) is reduced by
ca. 0.4 mag (instead of the usual limit of ca. 6.0 mag at zenith) due to the fact that the human eye has its 
sensitivity maximum in the green-yellow wavelength range. Since Mira culminates at
an altitude of $\sim 31.5^{\circ}$ at Fabricius' location, the increased atmospheric extinction and sky 
surface brightness at this zenith distance reduce the limit by further $\sim 0.6$ mag (Sect. 4.4).
\item We present multiple (additional) evidence that very red stars are hardly seen by the unaided eye when fainter than V$\simeq$4.5 mag
(Sect. 4.4).
\item According to the critically compiled data from Hevelius and Argelander, 
Mira can be observed by the unaided eye until 54-83 days after the maximum (Fabricius: since ca. 1596 Aug 22), 
so that Fabricius should have seen it until 1596 Oct 14 to Nov 12 --
indeed, he reported that Mira disappeared just around the time of a mentioned 
battle and conquest in Hungary, 1596 Oct 12-26 (Sects. 3.6 and 4.4).
\item The Mira brightening theory given by Fabricius, in connection to the prograde motion of the nearby Jupiter, explains why it took
him one Jupiter period of 12 years to detect it again (Sect. 3.6) -- so, he implicitly established
the existence of periodic variable stars to disappear and re-appear;
Fabricius: `the bodies of such new stars also later remain on sky, but invisible'.
\item We also discuss briefly Fabricius' record on observations of P Cygni in 1602, given as reddish and like Mars
(B$-$V=$1.43 \pm 0.13$ mag), narrated in connection to Mira; 
P Cygni was as bright as some `3rd mag' according to other observers (Sect. 3.4). 
\end{enumerate}

The solutions for the Mira discovery problem, in particular that it
was not reported as {\it new star} before 1596, are as follows: \\
(i) In medieval times and the Early Modern Period, 
mainstream Aristotelian cosmology and Christian/Jewish belief did not expect new stars.
Since Mira is located close to the ecliptic, it could have been mistaken for just another comet 
(see e.g. Fabricius in Sect. 3.2), where motion relative to stars and tail extension can be negligible,
if detected at sufficiently large separation. \\
(ii) Deep red stars like Mira are more difficult to detect by the naked eye (Sect. 4.4). \\
(iii) Why could Fabricius re-detect Mira only 12 years later? Fabricius combined the brightening of Mira
with the nearby prograde Jupiter (Sects. 3.6 \& 4.2). \\
(iv) East Asia: Mira may well have been a constituent of the Chinese constellation Tianjun (Sect. 2, part iv) --
and would therefore not be reported as {\it guest star} (Sect. 2, part iii). \\
(v) Mira is brighter than 3 mag only during about half of its maxima, then for only up to about one month, 
and about 3 out of 12 maxima (period 330.2 days) are when in conjunction with the Sun (April to June) --
hence, it is hard to discover serendipitously.\footnote{Even the (re-)detections of Mira after it was
discovered and known were difficult and very rare, namely after 1609 as follows: 
Wilhelm Schickard (1592-1635, studied at U T\"ubingen, professor for Hebrew and astronomy there,
inventor of a mechanical calculator) 
on 1631 Oct 14 at a maximum of 2.5 mag (see A.G. Pingré in his
unfinished work Histoire l'Astronomie du 17e siècle, p. 86);
Johannes Holwarda (1618-1651, professor for philosophy at Franeker, West Frisia, present-day The Netherlands) 
detected Mira serendipitously during a total lunar eclipse on 1638 Dec 24 ({\it jul.}) and again
in Nov 1639 (Hatch 2011, pp. 156-157), he noticed Mira's periodicity to be ca. 11 months; 
Bernard Fullenius (1602-1657, mathematician at Franeker) in 1639, 1641, and 1644;
Joachim Jungius (1587-1657, studied at U Padua, professor of mathematics at U Hamburg) in 1648 from Feb 18 to July; and
Johannes Hevelius (1611-1687, astronomer and mayor in Gdansk, Poland) with just one early sighting on 1648 Jan 5 
before the intensive monitoring by himself since 1659 and since 1663 also together with his wife Elisabeth nee Koopman (1647-1693)
as well as by Ismael Boulliau (1605-1694, catholic priest and astronomer in Paris) since 1661;
see Hatch (2011, following Hevelius 1662).}

Can one conclude on a limit (e.g. 3 mag) for serendipitous discoveries of `new stars' from the
Mira detection by Fabricius or the presumable non-detection by East Asian astronomers (see Introduction)? \\
Whether a new object is discovered serendipitously depends on the observing project (targets), 
as well as previous knowledge, carefulness, persistance, and biases of the observer(s) -- 
and also objective criteria (e.g. sky brightness of the given field and color index of that new star). There is no fixed limit.

\smallskip

\noindent {\bf Acknowledgements.}
RN and DLN have designed the study together and wrote most of the paper as interdisciplinary team.
MM concentrated on the naked-eye visibility dependence on color, DL on Latin texts, and JC on Chinese material.
All authors read and commented on the paper. \\
All data used in this work are either available within the article or in the references listed. \\
We acknowledge Frank Vohla for information on his naked-eye observations of Mira.
We made use of the data on AAVSO, in particular of Mira, and would like to thank all their observers. 
We used data from the Simbad and VizieR databases operated at the CDS in Strasbourg, France.
We also thank the involved libraries for the old manuscripts in digital form.
Wen-Chen Ping (Taiwan) has sent the article by Huang (1988), and
Frank Gie\ss ler (AIU, U Jena) has re-drawn of the upper right part of Fig. 1.
We acknowledge Brad Schaefer for discussion on sky brightness.
We thank the referee for critical questions and good advise.

{}

\section{Appendix: Chinese text}

Here, we provide the Chinese text from Sect. 2 (iii) in Chinese characters
together with our romanization and translation to English.

\begin{figure*}
\centering
\includegraphics[angle=0,width=18cm]{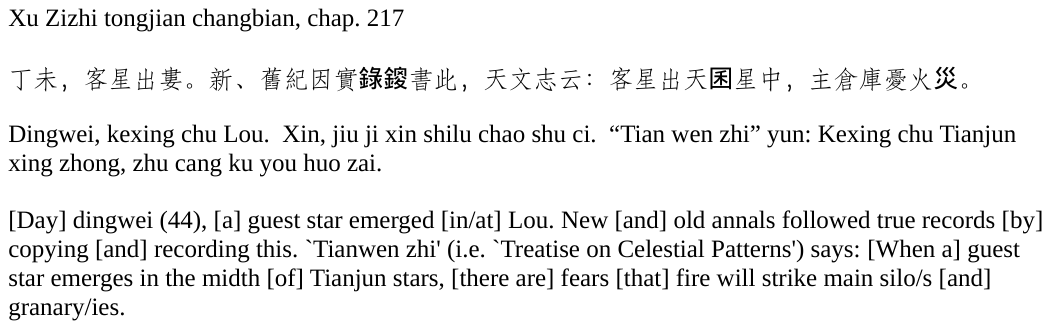}
\end{figure*}


\begin{thebibliography}{}
\bibitem[()]{} Caesius, G., 1579b, Chronick Oder ordenliche verzeichnu\ss~vnnd beschreibung aller Cometen von der algemeinen S\"undflut 
an bi\ss~auff dis gegenwertiges jtztlauffends 1579. Jar vnd was darauff f\"ur zufell straffen vnd verenderungen erfolget, Nuremberg
\end{thebibliography}
\end{document}